\documentclass[11pt]{article}
\usepackage{hyperref}
\usepackage{latexsym,graphicx,mathrsfs,amsfonts}
\usepackage{slashed, color}
\usepackage{multirow}
\usepackage{tikz}
\usepackage{amsmath}
\allowdisplaybreaks[2]
\usepackage{color}
\usepackage[multiple]{footmisc}
\usepackage{textcomp}
\usepackage{amssymb}
\usetikzlibrary{positioning,matrix,arrows,decorations.pathmorphing,decorations.pathreplacing}
\usepackage{amscd}
\usepackage{amsthm}
\usepackage{array} 

\renewcommand{\thefootnote}{\fnsymbol{footnote}}
\numberwithin{equation}{section}
\usepackage{dsfont}
\usepackage{booktabs}

\DeclareFontFamily{U}{MnSymbolC}{}
\DeclareSymbolFont{MnSyC}{U}{MnSymbolC}{m}{n}
\DeclareFontShape{U}{MnSymbolC}{m}{n}{
	<-6>  MnSymbolC5
	<6-7>  MnSymbolC6
	<7-8>  MnSymbolC7
	<8-9>  MnSymbolC8
	<9-10> MnSymbolC9
	<10-12> MnSymbolC10
	<12->   MnSymbolC12}{}
\DeclareMathSymbol{\intprod}{\mathbin}{MnSyC}{'270}
\newcommand{\ov}{\overline}
\newcommand{\C}{\mathbb{C}}

\newcommand{\Z}{\mathbb{Z}}
\newcommand{\R}{\mathbb{R}}

\newcommand{\del}{\partial}

\newcommand{\til}{\widetilde}

\let\nc\newcommand
\let\renc\renewcommand
\nc{\wbar}{\overline}
\let\td\tilde
\let\wtd\widetilde
\let\wht\widehat
\let\mcl\mathcal

\nc{\ab}{{\bar{a}}} \nc{\at}{\tilde{a}} \nc{\ah}{\hat{a}}
\nc{\bb}{{\bar{b}}} 
\nc{\bh}{\hat{b}}
\nc{\cb}{{\bar{c}}} \nc{\ct}{\tilde{c}} 
\nc{\db}{{\bar{d}}} \nc{\dt}{\tilde{d}} \renc{\dh}{\hat{d}}
\nc{\eb}{{\bar{e}}} \nc{\et}{\tilde{e}} \nc{\eh}{\hat{e}}
\nc{\fb}{{\bar{f}}} \nc{\ft}{\tilde{f}} \nc{\fh}{\hat{f}}
\nc{\gb}{{\bar{g}}} \nc{\gt}{\tilde{g}} \nc{\gh}{\hat{g}}
\nc{\ib}{{\bar{\imath}}} \nc{\ih}{\hat{\imath}} 
\nc{\jb}{{\bar{\jmath}}} \nc{\jt}{\tilde{\jmath}} \nc{\jh}{\hat{\jmath}}
\nc{\kb}{{\bar{k}}} \nc{\kt}{\tilde{k}} \nc{\kh}{\hat{k}}
\nc{\lb}{{\bar{l}}} \nc{\lt}{\tilde{l}} \nc{\lh}{\hat{l}}
\nc{\mb}{{\bar{m}}} \nc{\mt}{\tilde{m}} \nc{\mh}{\hat{m}}
\nc{\nb}{{\bar{n}}} \nc{\nt}{\tilde{n}} \nc{\nh}{\hat{n}}
\nc{\ob}{{\bar{o}}} \nc{\ot}{\tilde{o}} \nc{\oh}{\hat{o}}
\nc{\pb}{{\bar{p}}} \nc{\pt}{\tilde{p}} \nc{\ph}{\hat{p}}
\nc{\qb}{{\bar{q}}} \nc{\qt}{\tilde{q}} \nc{\qh}{\hat{q}}
\nc{\rb}{{\bar{r}}} \nc{\rt}{\tilde{r}} \nc{\rh}{\hat{r}}
\renc{\sb}{{\bar{s}}} \nc{\st}{\tilde{s}} \nc{\sh}{\hat{s}}
\nc{\tb}{{\bar{t}}} \renc{\th}{\hat{t}} 
\nc{\ub}{{\bar{u}}} \nc{\ut}{\tilde{u}} \nc{\uh}{\hat{u}}
\nc{\vb}{{\bar{v}}} \nc{\vt}{\tilde{v}} \nc{\vh}{\hat{v}}
\nc{\wt}{\tilde{w}} \nc{\wh}{\hat{w}}
\nc{\xb}{{\bar{x}}} \nc{\xt}{\tilde{x}} \nc{\xh}{\hat{x}}
\nc{\yb}{{\bar{y}}} \nc{\yt}{\tilde{y}} \nc{\yh}{\hat{y}}
\nc{\zb}{{\bar{z}}} \nc{\zt}{\tilde{z}} 

\nc{\Ab}{\wbar{A}} \nc{\At}{\wtd{A}} \nc{\Ah}{\wht{A}}
\nc{\Bb}{\wbar{B}} \nc{\Bt}{\wtd{B}} \nc{\Bh}{\wht{B}}
\nc{\Cb}{\wbar{C}} \nc{\Ct}{\wtd{C}} \nc{\Ch}{\wht{C}}
\nc{\Db}{\wbar{D}} \nc{\Dt}{\wtd{D}} \nc{\Dh}{\wht{D}}
\nc{\Eb}{\wbar{E}} \nc{\Et}{\wtd{E}} \nc{\Eh}{\wht{E}}
\nc{\Fb}{\wbar{F}} \nc{\Ft}{\wtd{F}} \nc{\Fh}{\wht{F}}
\nc{\Gb}{\wbar{G}} \nc{\Gt}{\wtd{G}} \nc{\Gh}{\wht{G}}
\nc{\Hb}{\wbar{H}} \nc{\Ht}{\wtd{H}} \nc{\Hh}{\wht{H}}
\nc{\Ib}{\wbar{I}} \nc{\It}{\wtd{I}} \nc{\Ih}{\wht{I}}
\nc{\Jb}{\wbar{J}} \nc{\Jt}{\wtd{J}} \nc{\Jh}{\wht{J}}
\nc{\Kb}{\wbar{K}} \nc{\Kt}{\wtd{K}} \nc{\Kh}{\wht{K}}
\nc{\Lb}{\wbar{L}} \nc{\Lt}{\wtd{L}} \nc{\Lh}{\wht{L}}
\nc{\Mb}{\wbar{M}} \nc{\Mt}{\wtd{M}} \nc{\Mh}{\wht{M}}
\nc{\Nb}{\wbar{N}} \nc{\Nt}{\wtd{N}} \nc{\Nh}{\wht{N}}
\nc{\Ob}{\wbar{O}} \nc{\Ot}{\wtd{O}} \nc{\Oh}{\wht{O}}
\nc{\Pb}{\wbar{P}} \nc{\Pt}{\wtd{P}} \nc{\Ph}{\wht{P}}
\nc{\Qb}{\wbar{Q}} \nc{\Qt}{\wtd{Q}} \nc{\Qh}{\wht{Q}}
\nc{\Rb}{\wbar{R}} \nc{\Rt}{\wtd{R}} \nc{\Rh}{\wht{R}}
\nc{\Sb}{\wbar{S}} \nc{\St}{\wtd{S}} \nc{\Sh}{\wht{S}}
\nc{\Tb}{\wbar{T}} \nc{\Tt}{\wtd{T}} \nc{\Th}{\wht{T}}
\nc{\Ub}{\wbar{U}} \nc{\Ut}{\wtd{U}} \nc{\Uh}{\wht{U}}
\nc{\Vb}{\wbar{V}} \nc{\Vt}{\wtd{V}} \nc{\Vh}{\wht{V}}
\nc{\Wb}{\wbar{W}} \nc{\Wt}{\wtd{W}} \nc{\Wh}{\wht{W}}
\nc{\Xb}{\wbar{X}} \nc{\Xt}{\wtd{X}} \nc{\Xh}{\wht{X}}
\nc{\Yb}{\wbar{Y}} \nc{\Yt}{\wtd{Y}} \nc{\Yh}{\wht{Y}}
\nc{\Zb}{\wbar{Z}} \nc{\Zt}{\wtd{Z}} \nc{\Zh}{\wht{Z}}

\nc{\CA}{\mcl{A}} \nc{\CAb}{\wbar{\CA}} \nc{\CAt}{\wtd{\CA}} \nc{\CAh}{\wht{\CA}}
\nc{\CB}{\mcl{B}} \nc{\CBb}{\wbar{\CB}} \nc{\CBt}{\wtd{\CB}} \nc{\CBh}{\wht{\CB}}
\nc{\CC}{\mcl{C}} \nc{\CCb}{\wbar{\CC}} \nc{\CCt}{\wtd{\CC}} \nc{\CCh}{\wht{\CC}}
\nc{\cDt}{\wtd{\cC}} \nc{\cDh}{\wht{\cD}}
\nc{\CE}{\mcl{E}} \nc{\CEb}{\wbar{\CE}} \nc{\CEt}{\wtd{\CE}} \nc{\CEh}{\wht{\CE}}
\nc{\CF}{\mcl{F}} \nc{\CFb}{\wbar{\CF}} \nc{\CFt}{\wtd{\CF}} \nc{\CFh}{\wht{\CF}}
\nc{\CG}{\mcl{G}} \nc{\CGb}{\wbar{\CG}} \nc{\CGt}{\wtd{\CG}} \nc{\CGh}{\wht{\CG}}
\nc{\CH}{\mcl{H}} \nc{\CHb}{\wbar{\CH}} \nc{\CHt}{\wtd{\CH}} \nc{\CHh}{\wht{\CH}}
\nc{\CI}{\mcl{I}} \nc{\CIb}{\wbar{\CI}} \nc{\CIt}{\wtd{\CI}} \nc{\CIh}{\wht{\CI}}
\nc{\CJ}{\mcl{J}} \nc{\CJb}{\wbar{\CJ}} \nc{\CJt}{\wtd{\CJ}} \nc{\CJh}{\wht{\CJ}}
\nc{\CK}{\mcl{K}} \nc{\CKb}{\wbar{\CK}} \nc{\CKt}{\wtd{\CK}} \nc{\CKh}{\wht{\CK}}
\nc{\CL}{\mcl{L}} \nc{\CLb}{\wbar{\CL}} \nc{\CLt}{\wtd{\CL}} \nc{\CLh}{\wht{\CL}}
\nc{\CM}{\mcl{M}} \nc{\CMb}{\wbar{\CM}} \nc{\CMt}{\wtd{\CM}} \nc{\CMh}{\wht{\CM}}
\nc{\CN}{\mcl{N}} \nc{\CNb}{\wbar{\CN}} \nc{\CNt}{\wtd{\CN}} \nc{\CNh}{\wht{\CN}}
\nc{\CO}{\mcl{O}} \nc{\COb}{\wbar{\CO}} \nc{\COt}{\wtd{\CO}} \nc{\COh}{\wht{\CO}}
\nc{\CQ}{\mcl{Q}} \nc{\CQb}{\wbar{\CQ}} \nc{\CQt}{\wtd{\CQ}} \nc{\CQh}{\wht{\CQ}}
\nc{\CR}{\mcl{R}} \nc{\CRb}{\wbar{\CR}} \nc{\CRt}{\wtd{\CR}} \nc{\CRh}{\wht{\CR}}
\nc{\CS}{\mcl{S}} \nc{\CSb}{\wbar{\CS}} \nc{\CSt}{\wtd{\CS}} \nc{\CSh}{\wht{\CS}}
\nc{\CT}{\mcl{T}} \nc{\CTb}{\wbar{\CT}} \nc{\CTt}{\wtd{\CT}} \nc{\CTh}{\wht{\CT}}
\nc{\CU}{\mcl{U}} \nc{\CUb}{\wbar{\CU}} \nc{\CUt}{\wtd{\CU}} \nc{\CUh}{\wht{\CU}}
\nc{\CV}{\mcl{V}} \nc{\CVb}{\wbar{\CV}} \nc{\CVt}{\wtd{\CV}} \nc{\CVh}{\wht{\CV}}
\nc{\CW}{\mcl{W}} \nc{\CWb}{\wbar{\CW}} \nc{\CWt}{\wtd{\CW}} \nc{\CWh}{\wht{\CW}}
\nc{\CX}{\mcl{X}} \nc{\CXb}{\wbar{\CX}} \nc{\CXt}{\wtd{\CX}} \nc{\CXh}{\wht{\CX}}
\nc{\CY}{\mcl{Y}} \nc{\CYb}{\wbar{\CY}} \nc{\CYt}{\wtd{\CY}} \nc{\CYh}{\wht{\CY}}
\nc{\CZ}{\mcl{Z}} \nc{\CZb}{\wbar{\CZ}} \nc{\CZt}{\wtd{\CZ}} \nc{\CZh}{\wht{\CZ}}

\let\eps\epsilon
\let\ups\upsilon
\let\veps\varepsilon
\let\vtht\vartheta
\let\vsgm\varsigma
\let\vphi\varphi
\let\vrho\varrho

\nc{\alphab}{\bar{\alpha}} \nc{\alphat}{\td{\alpha}} \nc{\alphah}{\hat{\alpha}}
\nc{\betab}{\bar{\beta}}   \nc{\betat}{\td{\beta}}   \nc{\betah}{\hat{\beta}} 
\nc{\gammab}{\bar{\gamma}} \nc{\gammat}{\td{\gamma}} \nc{\gammah}{\hat{\gamma}} 
\nc{\deltab}{\bar{\delta}} \nc{\deltat}{\td{\delta}} \nc{\deltah}{\hat{\delta}} 
\nc{\epsilonb}{\bar{\eps}} \nc{\epsilont}{\td{\eps}} \nc{\epsilonh}{\hat{\eps}} 
\nc{\vepsb}{\bar{\veps}}   \nc{\vepst}{\td{\veps}}   \nc{\vepsh}{\hat{\veps}} 
\nc{\zetab}{\bar{\zeta}}   \nc{\zetat}{\td{\zeta}}   \nc{\zetah}{\hat{\zeta}} 
\nc{\etab}{\bar{\eta}}     
\nc{\etah}{\hat{\eta}} 
\nc{\thetab}{\bar{\theta}} \nc{\thetat}{\td{\theta}} \nc{\thetah}{\hat{\theta}} 
\nc{\vthetab}{\bar{\vtht}} \nc{\vthetat}{\td{\vtht}} \nc{\vthetah}{\hat{\vtht}} 
\nc{\lambdat}{\td{\lambda}} \nc{\lambdah}{\hat{\lambda}} 
\nc{\iotab}{\bar{\iota}}   \nc{\iotat}{\td{\iota}}   \nc{\iotah}{\hat{\iota}} 
\nc{\kappab}{\bar{\kappa}} \nc{\kappat}{\td{\kappa}} \nc{\kappah}{\hat{\kappa}} 
\nc{\lmdb}{\bar{\lmd}}     \nc{\lmdt}{\td{\lmd}}     \nc{\lmdh}{\hat{\lmd}} 
\nc{\mub}{\bar{\mu}}       \nc{\mut}{\td{\mu}}       \nc{\muh}{\hat{\mu}} 
\nc{\nub}{\bar{\nu}}       \nc{\nut}{\td{\nu}}       \nc{\nuh}{\hat{\nu}} 
\nc{\xib}{\bar{\xi}}       \nc{\xit}{\td{\xi}}       \nc{\xih}{\hat{\xi}} 
\nc{\pib}{\bar{\pi}}       \nc{\pit}{\td{\pi}}       \nc{\pih}{\hat{\pi}} 
\nc{\vpib}{\bar{\vpi}}     \nc{\vpit}{\td{\vpi}}     \nc{\vpih}{\hat{\vpi}} 
\nc{\rhob}{\bar{\rho}}     \nc{\rhot}{\td{\rho}}     \nc{\rhoh}{\hat{\rho}} 
\nc{\vrhob}{\bar{\vrho}}   \nc{\vrhot}{\td{\vrho}}   \nc{\vrhoh}{\hat{\vrho}} 
\nc{\sigmab}{\bar{\sigma}} \nc{\sigmat}{\td{\sigma}} \nc{\sigmah}{\hat{\sigma}} 
\nc{\vsigmab}{\bar{\vsgm}} \nc{\vsigmat}{\td{\vsgm}} \nc{\vsigmah}{\hat{\vsgm}} 
\nc{\taub}{\bar{\tau}}     \nc{\taut}{\td{\tau}}     \nc{\tauh}{\hat{\tau}} 
\nc{\upsb}{\bar{\ups}} \nc{\upst}{\td{\ups}} \nc{\upsh}{\hat{\ups}} 
\nc{\phib}{\bar{\phi}}     \nc{\phit}{\td{\phi}}     \nc{\phih}{\hat{\phi}} 
\nc{\varphib}{\bar{\vphi}}   \nc{\varphit}{\td{\vphi}}   \nc{\varphih}{\hat{\vphi}} 
\nc{\chib}{\bar{\chi}}     
\nc{\chih}{\hat{\chi}} 
\nc{\psib}{\bar{\psi}}     
\nc{\psih}{\hat{\psi}} 
\nc{\omegab}{\bar{\omega}} \nc{\omegat}{\td{\omega}} \nc{\omegah}{\hat{\omega}} 

\nc{\Gammab}{\wbar{\Gamma}}     \nc{\Gammat}{\wtd{\Gamma}}     \nc{\Gammah}{\wht{\Gamma}}
\nc{\Deltab}{\wbar{\Delta}}     \nc{\Deltat}{\wtd{\Delta}}     \nc{\Deltah}{\wht{\Delta}}
\nc{\Thetab}{\wbar{\Theta}}     \nc{\Thetat}{\wtd{\Theta}}     \nc{\Thetah}{\wht{\Theta}}
\nc{\Lambdab}{\wbar{\Lambda}}   \nc{\Lambdat}{\wtd{\Lambda}}   \nc{\Lambdah}{\wht{\Lambda}}
\nc{\Xib}{\wbar{\Xi}}           \nc{\Xit}{\wtd{\Xi}}           \nc{\Xih}{\wht{\Xi}}
\nc{\Pib}{\wbar{\Pi}}           \nc{\Pit}{\wtd{\Pi}}           \nc{\Pih}{\wht{\Pi}}
\nc{\Sigmab}{\wbar{\Sigma}}     \nc{\Sigmat}{\wtd{\Sigma}}     \nc{\Sigmah}{\wht{\Sigma}}
\nc{\Upsilonb}{\wbar{\Upsilon}} \nc{\Upsilont}{\wtd{\Upsilon}} \nc{\Upsilonh}{\wht{\Upsilon}}
\nc{\Phib}{\wbar{\Phi}}         \nc{\Phit}{\wtd{\Phi}}         \nc{\Phih}{\wht{\Phi}}
\nc{\Psib}{\wbar{\Psi}}         \nc{\Psit}{\wtd{\Psi}}         \nc{\Psih}{\wht{\Psi}}
\nc{\Omegab}{\wbar{\Omega}}     \nc{\Omegat}{\wtd{\Omega}}     \nc{\Omegah}{\wht{\Omega}}

\newcommand{\cA}{{\cal A}}
\newcommand{\cAb}{{\overline{\cal A}}}

\newcommand{\Tr}{{\textrm{Tr}\;}}

\newcommand{\cF}{{\cal F}}
\newcommand{\cFb}{{\overline{\cal F}}}

\newcommand{\cD}{{\cal D}}
\newcommand{\cDb}{{\overline{\cal D}}}

\newcommand{\hlf}{\frac{1}{2}}

\newcommand{\si}{\sigma}

\newcommand{\hM}{\widehat{M}}
\newcommand{\hN}{\widehat{N}}

\newcommand{\ha}{\widehat{A}}
\newcommand{\hb}{\widehat{B}}
\newcommand{\hc}{\widehat{C}}
\newcommand{\x}{$\times$}
\nc{\balpha}{\bar{\alpha}}
\nc{\bbeta}{\bar{\beta}}
\nc{\bgamma}{\bar{\gamma}}
\nc{\bm}{\bar{m}}
\nc{\bn}{\bar{n}}
\nc{\bp}{\bar{p}}
\nc{\al}{\alpha}
\nc{\bt}{\beta}
\nc{\gm}{\gamma}
\nc{\zh}{\wht{z}}
\nc{\zhb}{\ov{\wht{z}}}
\nc{\mbh}{\wht{\ov{m}}}
\nc{\bc}{|_{x^3=0}}

\nc{\tal}{\til{\al}}
\nc{\tbt}{\til{\bt}}
\nc{\tgm}{\til{\gm}}

\nc{\wb}{\ov{w}}
\nc{\teta}{\til{\eta}}
\nc{\tpsi}{\til{\psi}}

\def\IL{\relax{\rm I\kern-.18em L}}
\def\IH{\relax{\rm I\kern-.18em H}}
\def\IB{\relax{\rm I\kern-.18em B}}
\def\ID{\relax{\rm I\kern-.18em D}}
\def\IE{\relax{\rm I\kern-.18em E}}
\def\IF{\relax{\rm I\kern-.18em F}}

\def\IG{\relax\hbox{$\inbar\kern-.3em{\rm G}$}}
\def\IGa{\relax\hbox{${\rm I}\kern-.18em\Gamma$}}
\def\IH{\relax{\rm I\kern-.18em H}}
\def\II{\relax{\rm I\kern-.18em I}}
\def\IK{\relax{\rm I\kern-.18em K}}
\def\IP{\relax{\rm I\kern-.18em P}}
\def\IQ{\relax\hbox{$\inbar\kern-.3em{\rm Q}$}}

\def\hat{\widehat}
\def\CM {{\cal M}}
\def\CN {{\cal N}}
\def\CR {{\cal R}}

\def\CF {{\cal F}}
\def\CJ {{\cal J}}

\def\CL {{\cal L}}
\def\CV {{\cal V}}
\def\CO {{\cal O}}
\def\CZ {{\cal Z}}
\def\CE {{\cal E}}
\def\CG {{\cal G}}
\def\CH {{\cal H}}
\def\CC {{\cal C}}
\def\CB {{\cal B}}
\def\CS {{\cal S}}
\def\CA{{\cal A}}
\def\CK{{\cal K}}
\def\CQ{{\cal Q}}

\def\p{\partial}
\def\pb{{\bar \p}}

\def\vt#1#2#3{ {\vartheta[{#1 \atop  #2}](#3\vert \tau)} }

\def\jb{{\bar j}}

\def\inbar{\,\vrule height1.5ex width.4pt depth0pt}

\def\half{{1 \over 2}}

\newcommand{\cbrac}[1]{\left(#1\right)}
\newcommand{\sbrac}[1]{\left[#1\right]}

\newcommand{\mbrac}[1]{\left.\left[#1\right.\right\}}
\newcommand{\tr}{\text{Tr}}

\newcommand{\susy}[1]{$\mathcal{N}=#1$}

\newcommand{\qcharge}{\mathcal{Q}}

\newcommand{\SO}[1]{SO\cbrac{#1}}

\newcommand{\real}{\mathbb{R}}
\newcommand{\complex}{\mathbb{C}}

\newcommand{\cM}{\mathcal{M}}
\newcommand{\cN}{\mathcal{N}}
\newcommand{\cQ}{\mathcal{Q}}
\newcommand{\cV}{\mathcal{V}}

\newcommand{\ovj}{\ov{\jmath}}
\newcommand{\tsi}{\til{\si}}
\newcommand{\tphi}{\til{\phi}}

\newcommand{\cB}{\mathcal{B}}

\newcommand{\cT}{\mathcal{T}}
\newcommand{\cS}{\mathcal{S}}

\let\OLDthebibliography\thebibliography
\renewcommand\thebibliography[1]{
	\OLDthebibliography{#1}
	\setlength{\parskip}{5pt}
	\setlength{\itemsep}{0pt plus 0.3ex}
}
\usepackage{titlesec}
\titleformat*{\section}{\bfseries\large}
\begin{document}
	\addtolength{\baselineskip}{1.5mm}
	
	\thispagestyle{empty}
	\vbox{}
	\vspace{3.0cm}
	
	\begin{center}
		\centerline{\LARGE{\textbf{Unifying Lattice Models, Links and Quantum}}}
		\medskip
		\centerline{\LARGE{\textbf{Geometric Langlands via Branes in String Theory}}}      
		
		\vspace{3.0cm}
		
		{Meer~Ashwinkumar\footnote{E-mail: meerashwinkumar@u.nus.edu} and Meng-Chwan~Tan\footnote{E-mail: mctan@nus.edu.sg}} 
		\\[2mm]
		{\it Department of Physics\\
			National University of Singapore \\
			2 Science Drive 3, Singapore 117551} \\[1mm] 
	\end{center}
	
	\vspace{2.0cm}

	\centerline{\bf Abstract}\smallskip \noindent
		We explain how, starting with a stack of D4-branes ending on an NS5-brane 
		in type IIA string theory, 
		one can, via T-duality and the topological-holomorphic nature of the relevant worldvolume theories, relate (i) the lattice models realized by Costello's 4d Chern-Simons theory, (ii) links in 3d analytically-continued Chern-Simons theory, (iii) the quantum geometric Langlands correspondence realized by Kapustin-Witten using 4d $\cN=4$ gauge theory and its quantum group modification, and (iv) the Gaitsgory-Lurie conjecture relating quantum groups/affine Kac-Moody algebras to Whittaker D-modules/W-algebras. This furnishes, purely physically via branes in string theory, a novel bridge between the mathematics of integrable systems, geometric topology, geometric representation theory, and quantum algebras.
	\newpage
	
	\renewcommand{\thefootnote}{\arabic{footnote}}
	\setcounter{footnote}{0}
	
	\tableofcontents
	\section{Introduction, Summary and Acknowledgements}
		\noindent{\textit{Introduction}}

	Chern-Simons theories in dimensions higher than three have recently been found to have intriguing mathematical and physical implications. In particular, 4d Chern-Simons theory studied by Costello, Witten and Yamazaki \cite{CWY,CWY2} realizes the Yang-Baxter equation with spectral parameter for various integrable lattice models, as well as the underlying Yangian algebra, quantum affine algebra, and elliptic quantum group for rational, trigonometric and elliptic solutions of the Yang-Baxter equation, respectively. 
	 Moreover, a non-commutative deformation of 5d Chern-Simons theory introduced by Costello has also been shown to encapsulate $\Omega$-deformed M-theory \cite{CostelloMtheory,GaioOh}. 
	
	In fact, 4d and 5d Chern-Simons theory form a family together with 3d analytically-continued Chern-Simons theory and 6d holomorphic Chern-Simons theory; they are related by successive dimensional reductions from the 6d theory, and the theories are moreover T-dual as QFTs \cite{yamazaki}. 
	This suggests that mathematical quantities realized in these various Chern-Simons theories should be related in suitable limits. It is well-known, for example, that knot invariants are obtainable from suitable limits of integrable models, with the braid group relation descending from the Yang-Baxter equation \cite{Wadati}.
	
	In another direction, 3d analytically-continued Chern-Simons theory 
	should be related to a generalization of the geometric Langlands correspondence, namely, the quantum geometric Langlands correspondence, 
	as the latter has an extension for quantum groups known as the Gaitsgory-Lurie conjecture
	\cite{Gaitsgory}.
	Moreover, it was shown by Witten \cite{witten2011fivebranes} that the four-dimensional topological field theory that realizes geometric Langlands \cite{kapustin2006electric} realizes 3d analytically-continued Chern-Simons theory on the boundary of the underlying four-manifold. This was an essential step in the 5d gauge theory approach to Khovanov homology, which itself has been defined using moduli spaces of geometric Hecke transformations \cite{CKKnot}, an essential ingredient in geometric Langlands.

In this paper, we will show, in the framework of string theory, that one can indeed relate lattice models realized by 4d Chern-Simons theory to links in 3d analytically-continued Chern-Simons theory, and consequently to the quantum geometric Langlands correspondence (that relates categories of twisted D-modules) and a quantum group modification thereof, 
and the Gaitsgory-Lurie conjecture, thereby furnishing a novel bridge between the mathematics of integrable systems, geometric topology, geometric representation theory, and quantum algebras.
\newline
\newline
	\noindent{\textit{Summary}}

 To achieve this, we begin by using a partial twist of 5d $\cN=2$ supersymmetric Yang-Mills (SYM) theory developed in \cite{AshTanQin} by the authors. We show that this twist gives rise to a 5d analogue of the geometric Langlands twist of 4d $\cN=4$ SYM, that also leads to a family of twisted theories parametrized by a complex variable, $t$. Such a partially-twisted theory can be understood to be the worldvolume theory of a stack of D4-branes in a non-trivial type IIA string theory background, where boundary conditions corresponding to a (deformed) NS5-brane 
 are also imposed. The 5d partially-twisted action on a five-manifold, $\cM$, is shown to take the form
\begin{equation}\label{inttopholo}
\boxed{
S=\{\cQ,\til{V}\} +{w-\bar{w} \over 4}{i\Psit\over 2\pi}\int_{\partial \cM}\ dz_w\wedge\tr\cbrac{\cA_w\wedge d\cA_w+{2\over 3}\cA_w\wedge \cA_w\wedge \cA_w},
}\end{equation}	
for the relevant supercharge, $\cQ$, and parameters $\Psit$ and $w$ .
 
 As we shall explain, the path integral of the 5d partially-twisted theory obtained from \eqref{inttopholo} localizes to 4d Chern-Simons theory 
\begin{equation}\label{intro4dcsf}
\boxed{
\int_{\til{\Gamma}} D\cA~\textrm{exp}\Bigg({\Psit\textrm{Im}(w)\over 4\pi}\int_{\partial \cM}\ dz\wedge\tr\cbrac{\cA\wedge d\cA+{2\over 3}\cA\wedge \cA\wedge \cA}\Bigg),}
\end{equation}	
on the boundary, where $\til{\Gamma}$ denotes an appropriate integration cycle defined by gradient flow equations 
\begin{equation}\label{intrograd1}
\boxed{
\begin{aligned}
\cF_{3\til{\gamma}}&=\pm i 2\veps_{\til{\gamma}}^{~\til{\alpha}}\ov{\cF}_{\til{\alpha}z}\\
	\cF_{3\zb}&=\pm\frac{i}{4}\veps^{\til{\beta}\til{\gamma}}\ov{\cF}_{\til{\beta}\til{\gamma}},
\end{aligned}}
\end{equation}
 and a supplementary constraint. This integration cycle allows 4d Chern-Simons theory to be defined beyond perturbation theory \cite{Witten}.
 
 Then, taking T-duality of this D4-NS5 system, we arrive at the D3-NS5 system studied by Witten in \cite{witten2011fivebranes} that realizes 3d analytically-continued Chern-Simons theory. Here, lattices in the former (realized by fundamental strings) are identified with links in the latter. We then explain how these T-dual systems can be modified to realize quantum geometric Langlands duality, by the inclusion of an additional D-brane at the other end of the D4-NS5 system such that T-duality give us a D3-NS5-D5 system.
 This shall be shown, via a further duality, to correspond in the low-energy limit to branes of the 2d sigma model on Hitchin's moduli space, $\cM_H(G,C)$, that realize twisted D-modules. Then, S-duality of type IIB string theory realizes quantum geometric Langlands duality that identifies twisted D-modules on $\cM_H(G,C)$ and $\cM_H(^LG,C)$, where $^LG$ is the Langlands dual of $G$. In fact, when the aforementioned links are in non-trivial representations of the relevant quantum 
 group,  we have a quantum group modification of this Langlands duality.
 
 In addition, we shall also show how S-duality of type IIB string theory 
 leads us to S-duality of 3d analytically-continued Chern-Simons theory previously uncovered in \cite{TY,DimofteGukov}. Moreover, we discuss an extension of quantum geometric Langlands duality realized in our setup, 
 namely, the aforementioned Gaitsgory-Lurie conjecture \cite{Gaitsgory} that relates the Kazhdan-Luzstig category of representations of quantum groups $U_q(G)$ - or equivalently, the category of finitely generated modules over the affine Kac-Moody algebra $\widehat{\mathfrak{g}}$ on which the action of $\mathfrak{g}[[z]]$ integrates to an action of the group $G[[z]]$ - to the category of Whittaker D-modules on the affine Grassmannian of $^LG$ - or equivalently to a category of modules of the affine W-algebra of $^LG$. Our findings are summarized in 
 Figure 1, which shall be elaborated on in Section 5.
  \begin{figure}
 \catcode`\@=11
\newdimen\cdsep
\cdsep=3em

\def\cdstrut{\vrule height .6\cdsep width 0pt depth .4\cdsep}
\def\@cdstrut{{\advance\cdsep by 2em\cdstrut}}

\def\arrow#1#2{
  \ifx d#1
    \llap{$\scriptstyle#2$}\left\downarrow\cdstrut\right.\@cdstrut\fi
  \ifx u#1
    \llap{$\scriptstyle#2$}\left\uparrow\cdstrut\right.\@cdstrut\fi
  \ifx r#1
    \mathop{\hbox to \cdsep{\rightarrowfill}}\limits^{#2}\fi
  \ifx l#1
    \mathop{\hbox to \cdsep{\leftarrowfill}}\limits^{#2}\fi
}
\catcode`\@=12

\cdsep=3em
$$
\begin{matrix}
 & & \textrm{KL}_{\Psi}(G)/\widehat{\mathfrak{g}}_{\Psi-h^{\vee}}\textrm{-}mod_C^0                & \arrow{r}{\cS'}   & \textrm{Whit}_{\frac{1}{\Psi}}(^LG)/\mathcal{W}_{\frac{1}{\Psi}}(^L\mathfrak{g})\textrm{-}mod^0_C                  \cr
 &&   \arrow{u}{\cT^{-1}} &                      & \arrow{u}{\cS'\cT^{-1}\cS'^{-1}} \cr
&  \textrm{4d CS}_{\Psit }(G_{\C})                    & \arrow{r}{\cT \textrm{T}}            \textrm{3d CS}_{\Psi +1}(G_{\C})  & \arrow{r}{\cS'}   & \textrm{3d CS}_{\frac{1}{\Psi +1} }(^LG_{\C})                    \cr
 && \arrow{d}{C\rightarrow 0} &                      & \arrow{d}{C\rightarrow 0} \cr
&&  D\textrm{-}mod^{U_{q'}(G_{\rho^i}),\rho^i}_{\Psi+1}(\cM(G,C))                 & \arrow{r}{\cS'} & D\textrm{-}mod^{U_{^Lq'}(^LG_{^L\rho^i}),^L\rho^i}_{\frac{1}{\Psi+1}}(\cM(^LG,C))                  \cr
\end{matrix}
$$ 
\caption{A relationship between 4d Chern-Simons theory, 3d S-dual Chern-Simons theories, the quantum group modification of quantum geometric Langlands, and the Gaitsgory-Lurie conjecture. Here, $q'=\textrm{exp}(\frac{\pi i}{\Psi+1})$.
}
\end{figure}
   This is how we relate lattice models to links and the quantum geometric Langlands correspondence, and its extensions, which realizes a novel bridge between the mathematics of integrable systems, geometric topology, geometric representation theory, and quantum algebras.
 \newline
 \newline
 	\noindent{\textit{Acknowledgements}}
 	
 	We would like to thank Masahito Yamazaki for helpful discussions. 
 	Results in 
 	this paper were presented by M. Ashwinkumar at Kavli IPMU in May, 2019, and by M.-C. Tan at the 11th International Symposium on Quantum Theory and Symmetries at the University of Montreal in July, 2019, and we thank the audiences of these talks for feedback. M. Ashwinkumar would also like to thank Kavli IPMU
 	for hospitality. This work is supported in part by the NUS FRC Tier 1 grant R-144-000-377-114 and the MOE Tier 2 grant R-144-000-396-112.
	\section{Partial Twist of 5d $\mathcal{N}=2$ Supersymmetric Yang-Mills}
\subsection{5d $\mathcal{N}=2$ Supersymmetric Yang-Mills}	
	We begin with the 5d maximally supersymmetric Yang-Mills theory (MSYM), obtained in the low-energy limit 
	of a stack of D4-branes in type IIA string theory. The classical action may be written as
\begin{equation}
\begin{aligned}
\label{untwisted}
S = -\frac{1}{g_5^2} \int_{\cM} d^5x ~\Tr \Big(&\hlf F_{MN} F^{MN} +  D_M \phi_{\hM} D^M \phi^{\hM} + \hlf [\phi_{\hM}, \phi_{\hN}] [\phi^{\hM}, \phi^{\hN}] 
-i{\rho}^{A{\ha}}(\Gamma^M)_A^{~~B} D_M \rho_{B\ha} \\&- {\rho}^{A\ha} (\Gamma^{\hM})_{\ha}^{~~\hb} [\phi_{\hM}, \rho_{A\hb}] \Big),
\end{aligned}
\end{equation} 
which is invariant under the supersymmetry transformations 
\begin{equation}\label{untwistedsusy}
\begin{aligned}
\delta A_M &= i\zeta ^{A\ha} (\Gamma_M)_A^{~~B}\rho_{B\ha}\\
\delta \phi ^{\hM}&=\zeta ^{A\ha} (\Gamma^{\hM})_{\ha}^{~~\hb}\rho_{A\hb}\\
\delta\rho _{A\ha}&=-i(\Gamma^M)_A^{~~B}D_M\phi^{\hM}(\Gamma_{\hM})_{\ha}^{~~\hb}\zeta_{B\hb}-\frac{1}{2}(\Gamma_{\hM})_{\ha}^{~~\hb}(\Gamma_{\hN})_{{\hb}\hc}[\phi^{\hM},\phi^{\hN}]\zeta_A^{~~\hc}+\frac{1}{2}F^{MN}(\Gamma_{MN})_{A}^{~~B}\zeta_{B{\ha}}.
\end{aligned}
\end{equation}
Here, $\cM$ in the subscript of the integral in \eqref{untwisted} denotes a flat 5d Euclidean manifold. Also, $(M,N,\ldots)$ and $(A,B,\ldots)$ are respectively vector and spinor indices for the $SO_{E}(5)$ rotation group, with their hatted counterparts corresponding to the $SO_R(5)$ R-symmetry group. In addition, the Lie algebra of the $U(N)$ gauge group is taken to be generated by antihermitian matrices $T_a$, where $a=1,\ldots,\textrm{dim }\mathfrak{u}(N)$, implying that the invariant quadratic form on this Lie algebra, denoted $\textrm{Tr}$, is negative-definite. In particular, the matrices $T_a$ are chosen such that $\textrm{Tr}(T_a T_b)=-\delta_{ab}$.
	
	The 5d action and supersymmetry transformations given here can be obtained from those of 10d \susy{1} supersymmetric Yang-Mills theory, i.e.,
\begin{equation}\label{10da}
S_{10}=-{1\over e^2}\int
d^{10}x\,\Tr\left({1\over
2}F_{IJ}F^{IJ}-i\bar\lambda\Gamma^ID_I\lambda\right)
\end{equation} 
and
\begin{equation}
\begin{aligned}
\delta A_I & = i\bar\epsilon\Gamma_I\lambda
\cr
                   \delta\lambda & = {1\over 2}
                   \Gamma^{IJ}F_{IJ},
                   \end{aligned}
\end{equation}	
via dimensional reduction. Here, we have used the notation and conventions of Kapustin and Witten \cite{kapustin2006electric}.
	\subsection{Partial Topological Twist}
	
	Consider next, a flat 5-manifold $\cM=Y\times\real_+\times\Sigma$, where $Y$ and $\Sigma$ are 2-manifolds corresponding to the $\{x^1,x^2\}$ and $\{x^4,x^5\}$ directions respectively, while $\R_+$ is half of the real line, $\R$, that corresponds to the $x^3$ direction. We shall twist along $V=Y\times \R_+$, by redefining its $SO_E(3)$ rotation group to be the diagonal subgroup $SO_E(3)'$ of $SO_E(3)\times SO_R(3)$, where $SO_R(3)$ is the subgroup of the R-symmetry group that rotates $\{\phi_{\wht{1}},\phi_{\wht{2}},\phi_{\wht{3}}\}$. 
	
	Put differently, we are considering the rotation subgroup $SO_E(3)\times SO_E(2)\subset SO_E(5)$, and also the R-symmetry subgroup $ SO_R(3)\times SO_R(2)\subset SO_R(5)$, and defining a new rotation group, $SO_E(3)'$, along $V=Y\times \R_+$.
	Before we proceed, we first specify our notational conventions for vector and spinor indices of the various symmetry groups. These are given in the following table: 
	\begin{center}
  \begin{tabular}{|l*{5}{c|}r}
    \hline
    & & $SO_V(3)$ & $SO_R(3)$ & $SO_{\Sigma}(2)$ & $SO_R(2)$ \\ \hline
    & Vector & $\alpha,\beta,\gamma,\ldots$ & $\wht{\alpha},\wht{\beta},\wht{\gamma},\ldots$ & $m,n,p,\ldots$ & $\wht{m},\wht{n},\wht{p},\ldots$\\ \hline
    & Spinor & $\bar{\alpha},\bar{\beta},\bar{\gamma},\ldots$ & $\wht{\bar{\alpha}},\wht{\bar{\beta}},\wht{\bar{\gamma}},\ldots$ & $\bar{m},\bar{n},\bar{p},\ldots$ & $\wht{\bar{m}},\wht{\bar{n}},\wht{\bar{p}},\ldots$  \\
    \hline
  \end{tabular}
\end{center}
Twisting amounts to setting the hatted $SO_R(3)$ indices to unhatted indices.

The twisting results in the scalar fields $\{\phi_{\wht{1}},\phi_{\wht{2}},\phi_{\wht{3}}\}$ now transforming as the components $\{\phi_{{1}},\phi_{{2}},\phi_{{3}}\}$ of a one-form on $Y\times \R_+$. Furthermore, the twisting of the fermions which transform as (\textbf{2},\textbf{2}) under $SO_E(3)\times SO_R(3)$ results in fermions which transform as \textbf{1} and \textbf{3}  under $SO_E(3)'$, i.e.,
\begin{equation}
\textbf{2} \otimes \textbf{2} = \textbf{1} \oplus \textbf{3}.
\end{equation}
We can see this explicitly by expanding the spinor fields $\rho_{A\wht{A}}=\rho_{\balpha \bm \wht{\balpha}\wht{\bm}}$, after twisting, as 
\begin{equation}\label{fermiexp}
\rho_{\balpha \bm {\bbeta}\wht{\bm}}= \eps_{\balpha \bbeta} \eta_{\bm \wht{\bm}} + (\si^\alpha)_{\balpha \bbeta} \psi_{\alpha\bm \wht{\bm}},
\end{equation}
where we have 
used the antisymmetric matrix $\epsilon_{\balpha \bbeta}$ and the symmetric matrix $(\si^\alpha)_{\balpha \bbeta}$ introduced in the appendix. The supersymmetry transformation parameters $\zeta_{A\wht{A}}=\zeta_{\balpha \bm \wht{\balpha}\wht{\bm}}$ may also be expanded in this manner, i.e., 
\begin{equation}\label{susyexp}
\zeta_{\balpha \bm {\bbeta}\wht{\bm}}= \eps_{\balpha \bbeta} \zeta_{\bm \wht{\bm}} + (\si^\alpha)_{\balpha \bbeta} \zeta_{\alpha\bm \wht{\bm}}.
\end{equation}
Using the explicit representation of the gamma matrices given in the appendix, we can substitute \eqref{fermiexp} and \eqref{susyexp} into \eqref{untwistedsusy} to obtain the partially-twisted supersymmetry transformations.

Now, we shall pick a supercharge, $\cQ$, that is scalar along $V$, with respect to which we shall eventually localize the theory.
The parameters $\zeta_{\bar{m}\hat{\bar{m}}}$ in (\ref{susyexp}) transform as scalars under $\SO{3}'$. 
We shall choose only two of these parameters to be non-zero, namely $\zeta_{11}$ and $\zeta_{21}$, and take a linear combination of their corresponding supercharges to be $\cQ$. The first reason for this choice is that it gives us supersymmetric localization equations that will eventually ensure the convergence of the 4d Chern-Simons path integral.
Secondly, this choice shall also lead to a relation to the geometric Langlands twist of 4d
	\susy{4} Yang-Mills theory \cite{kapustin2006electric}, in which the topological supercharge is a linear combination of two scalar supercharges. 
	Setting $\zeta_{11}=\kappa$ and $\zeta_{21}=\lambda$, where $\kappa,\lambda\in\complex$, the supersymmetry transformations are 
	
	\begin{equation}
	\label{eqn:5dtwistedsusy}
	\begin{aligned}
	\delta A_\alpha&=-2i\kappa\psi_{\alpha 22}+2i\lambda\psi_{\alpha 12}\\
	\delta \phi_\alpha&=2\kappa\psi_{\alpha 22}+2\lambda\psi_{\alpha 12}\\
	\delta A_4&=2i\kappa \eta_{12}+2i\lambda\eta_{22}\\
	\delta A_5&=-2\kappa\eta_{12}+2\lambda\eta_{22}\\
	\delta \phi_{\hat{4}}&=2\kappa\eta_{21}+2\lambda\eta_{11}\\
	\delta \phi_{\hat{5}}&=2i\kappa\eta_{21}+2i\lambda\eta_{11}\\
	\delta \psi_{\alpha 11}&=\kappa\varepsilon_{\alpha\beta\gamma}\cbrac{{i\over 2}F^{\beta\gamma}-{i\over 2}\sbrac{\phi^\beta,\phi^\gamma}-D^\beta\phi^\gamma}\\
	&\qquad+\lambda\cbrac{F_{\alpha 4}-iF_{\alpha 5}+i\cbrac{D_4-iD_5}\phi_\alpha}\\
	\delta\psi_{\alpha 12}&=\kappa\cbrac{\sbrac{\phi_\alpha,\phi_{\hat{4}}+i\phi_{\hat{5}}}-iD_\alpha\cbrac{\phi_{\hat{4}}+i\phi_{\hat{5}}}}\\
	\delta \psi_{\alpha 21}&=\kappa\cbrac{-F_{\alpha 4}-iF_{\alpha 5}+i\cbrac{D_4+iD_5}\phi_\alpha}\\
	&\qquad+\lambda\varepsilon_{\alpha\beta\gamma}\cbrac{{i\over 2}F^{\beta\gamma}-{i\over 2}\sbrac{\phi^\beta,\phi^\gamma}+D^\beta\phi^\gamma}\\
	\delta\psi_{\alpha 22}&=\lambda\cbrac{\sbrac{\phi_\alpha,\phi_{\hat{4}}+i\phi_{\hat{5}}}+iD_\alpha\cbrac{\phi_{\hat{4}}+i\phi_{\hat{5}}}}\\
	\delta\eta_{11}&=i\kappa\cbrac{F_{45}+\sbrac{\phi_{\hat{4}},\phi_{\hat{5}}}+D_\beta\phi^\beta}\\
	\delta\eta_{12}&=-i\lambda\cbrac{D_{4}-iD_5}\cbrac{\phi_{\hat{4}}+i\phi_{\hat{5}}}\\
	\delta\eta_{21}&=-i\lambda\cbrac{F_{45}-\sbrac{\phi_{\hat{4}},\phi_{\hat{5}}}+D_\beta\phi^\beta}\\
	\delta\eta_{22}&=-i\kappa\cbrac{D_4+iD_5}\cbrac{\phi_{\hat{4}}+i\phi_{\hat{5}}}.
	\end{aligned}
	\end{equation}
		\newline
	\newline
	\noindent{\textit{Field Redefinitions}}
	
	In what follows, it shall be convenient to redefine the scalars $\phi_{\hat{4}},\phi_{\hat{5}}$, fermi fields $\psi_\alpha,\eta$ and the supersymmetry parameters $\kappa,\lambda$, as it will allow us to identify them with objects in the 4d GL-twisted theory.
	 For the scalars, we have
	\begin{equation}
	\label{redef:scalar}
	\begin{aligned}
	\sigma={1\over \sqrt{2}}\cbrac{\phi_{\hat{5}}-i\phi_{\hat{4}}},&&\bar{\sigma}={1\over \sqrt{2}}\cbrac{\phi_{\hat{5}}+i\phi_{\hat{4}}}.
	\end{aligned}
	\end{equation}
	$\sigma$ here is a complex scalar field that is not to be confused with the Pauli matrices. For the fermions, we have
	\begin{equation}
	\label{redef:fermions}
	\begin{aligned}
	&\chi_\alpha={(1-i)\over 2^{5/4}}\psi_{\alpha 11}+{(-1-i)\over 2^{5/4}}\psi_{\alpha 21},&&\widetilde{\chi}_\alpha={(-1-i)\over 2^{5/4}}\psi_{\alpha 11}+{(1-i)\over 2^{5/4}}\psi_{\alpha 21}\\
	&\eta={(1+i)\over 2^{1/4}}\eta_{11}+{(1-i)\over 2^{1/4}}\eta_{21},&&\widetilde{\eta}={(-1+i)\over 2^{1/4}}\eta_{11}+{(-1-i)\over 2^{1/4}}\eta_{21}\\
	&\psi_\alpha={(1+i)\over 2^{3/4}}\psi_{\alpha 12}+{(-1+i)\over 2^{3/4}}\psi_{\alpha 22},&&\widetilde{\psi}_\alpha={(-1+i)\over 2^{3/4}}\psi_{\alpha 12}+{(1+i)\over 2^{3/4}}\psi_{\alpha 22}\\
	&\Upsilon={(1-i)\over 2^{3/4}}\eta_{12}+{(1+i)\over 2^{3/4}}\eta_{22},&&\widetilde{\Upsilon}={(-1-i)\over 2^{3/4}}\eta_{12}+{(-1+i)\over 2^{3/4}}\eta_{22}.
	\end{aligned}
	\end{equation}
	 Finally, the supersymmetry parameters are
	\begin{equation}
	\label{redef:susyparameters}
	\begin{aligned}
	u={1\over 2^{1/4}}\sbrac{(1+i)\kappa+(1-i)\lambda},&&v={1\over 2^{1/4}}\sbrac{(-1+i)\kappa+(-1-i)\lambda}.
	\end{aligned}
	\end{equation}
	
	The supersymmetry transformations in (\ref{eqn:5dtwistedsusy}) then become
	\begin{equation}
	\label{eqn:5dtwistedsusynew}
	\begin{aligned}
	\delta A_\alpha&=iu\psi_\alpha+iv\widetilde{\psi}_\alpha\\
	\delta \phi_\alpha&=iv\psi_\alpha-iu\widetilde{\psi}_\alpha\\
	\delta A_4&=iu\Upsilon+iv\widetilde{\Upsilon}\\
	\delta A_5&=iv\Upsilon-iu\widetilde{\Upsilon}\\
	\delta \sigma&=0\\
	\delta \bar{\sigma}&=iu\eta+iv\widetilde{\eta}\\
	\delta \chi_{\alpha}&=\half u\sbrac{F_{\alpha 4}+D_5\phi_\alpha+\half\varepsilon_{\alpha\beta\gamma}\cbrac{F^{\beta\gamma}-\sbrac{\phi^\beta,\phi^\gamma}}}\\
	&\qquad+\half v\sbrac{F_{\alpha 5}-D_4\phi_\alpha+\varepsilon_{\alpha\beta\gamma}D^\beta\phi^\gamma}\\
	\delta \widetilde{\chi}_{\alpha}&=\half v\sbrac{F_{\alpha 4}+D_5\phi_\alpha-\half\varepsilon_{\alpha\beta\gamma}\cbrac{F^{\beta\gamma}-\sbrac{\phi^\beta,\phi^\gamma}}}\\
	&\qquad-\half u\sbrac{F_{\alpha 5}-D_4\phi_\alpha-\varepsilon_{\alpha\beta\gamma}D^\beta\phi^\gamma}\\
	\delta\eta&=v\cbrac{F_{45}+D_\alpha\phi^\alpha}+u\sbrac{\bar{\sigma},\sigma}\\
	\delta\widetilde{\eta}&=-u\cbrac{F_{45}+D_\alpha\phi^\alpha}+v\sbrac{\bar{\sigma},\sigma}\\
	\delta\psi_\alpha&=uD_\alpha\sigma+v\sbrac{\phi_\alpha,\sigma}\\
	\delta\widetilde{\psi}_\alpha&=vD_\alpha\sigma-u\sbrac{\phi_\alpha,\sigma}\\
	\delta\Upsilon&=uD_4\sigma+vD_5\sigma\\
	\delta\widetilde{\Upsilon}&=vD_4\sigma-uD_5\sigma.
	\end{aligned}
	\end{equation}
We shall use this convenient form of the supersymmetry transformations in what follows.	
	
	\subsection{Relation to GL-twisted \susy{4} Super Yang-Mills}
	
	
	Now, we shall show that the supersymmetry transformations in \eqref{eqn:5dtwistedsusynew} are related to those of the GL-twisted theory in 4d which was studied in \cite{kapustin2006electric}. To be precise, the supersymmetry transformations in (\ref{eqn:5dtwistedsusynew}) reduce to those of the GL-twisted theory in 4d via dimensional reduction.
	
	To perform the dimensional reduction we shall take $\Sigma =\R\times S^1$, where $S^1$ is along the $x^5$ direction. We then dimensionally reduce along $S^1$. 
	Then, the 5th component of the gauge fields becomes disassociated, and we make the replacement $A_5\to\phi_4$. This is not to be confused with $\phi_{\hat{4}}$, which was used in the definition of $\sigma$ in (\ref{redef:scalar}). Instead, $\phi_4$ is to be viewed as the 4th component of the scalar fields -- i.e. $\phi_\mu=\cbrac{\phi_\alpha,\phi_4}$, where we have taken $\mu=1,2,3,4$. 
	Also, because there is no longer any dependence on the $x^5$-direction, derivatives in the compactified direction simply vanish. Consequently, we get the following reductions:
	\begin{equation}
	\begin{aligned}
	F_{\mu 5}&\to D_\mu\phi_4\\
	D_5&\to\sbrac{\phi_4,\ \cdot\ }.
	\end{aligned}
	\end{equation}

For the fermions, we first make the identification	$\chi_{\alpha}=\cbrac{\chi^+}_{\alpha 4}$ and $\widetilde{\chi}_{\alpha}=\cbrac{\chi^-}_{\alpha 4}$, where $\chi^\pm$ are 4d self-dual/anti-self-dual tensors that satisfy
	\begin{equation}\label{sasd}
	\cbrac{\chi^\pm}_{\mu\nu}=\pm\half\varepsilon_{\mu\nu}^{~~~\rho\sigma}\cbrac{\chi^\pm}_{\rho\sigma}.
	\end{equation}
	To obtain the $\alpha\beta$ components of $\chi^\pm$ we use
	\begin{equation}
	\begin{aligned}
	\label{con:antiselfdual}
	\cbrac{\chi^+}_{\alpha\beta}&=\half\varepsilon_{\alpha\beta}^{~~~\gamma }\chi_{\gm}\\
		\cbrac{\chi^-}_{\alpha\beta}&=-\half\varepsilon_{\alpha\beta}^{~~~\gamma }\til{\chi}_{\gm}.
	\end{aligned}
	\end{equation}	
since this identification implies
	\begin{equation}
	\label{con:antiselfdual}
	\cbrac{\chi^\pm}_{\alpha\beta}=\pm\half\varepsilon_{\alpha\beta}^{~~~\gamma 4}\cbrac{\chi^\pm}_{\gamma 4},
	\end{equation}
	in agreement with \eqref{sasd}.
Finally, we identify $\Upsilon$ and $\widetilde{\Upsilon}$ as $\psi_4$ and $\til{\psi}_4$, respectively.
	
	The supersymmetry transformations (\ref{eqn:5dtwistedsusynew}) then become
	\begin{equation}\label{newsus}
	\begin{aligned}
	\delta A_\mu&=iu\psi_\mu+iv\widetilde{\psi}_\mu\\
	\delta \phi_\mu&=iv\psi_\mu-iu\widetilde{\psi}_\mu\\
	\delta \sigma&=0\\
	\delta \bar{\sigma}&=iu\eta+iv\widetilde{\eta}\\
	\delta \cbrac{\chi^+}_{\mu\nu}&=u\cbrac{F_{\mu \nu}-\sbrac{\phi_\mu,\phi_\nu}}^+ +v\cbrac{D_\mu\phi_\nu}^+\\
	\delta \cbrac{\chi^-}_{\mu\nu}&=v\cbrac{F_{\mu\nu}-\sbrac{\phi_\mu,\phi_\nu}}^- -u\cbrac{D_\mu\phi_\nu}^-\\
	\delta\eta&=v\cbrac{D_\mu\phi^\mu}+u\sbrac{\bar{\sigma},\sigma}\\
	\delta\widetilde{\eta}&=-u\cbrac{D_\mu\phi^\mu}+v\sbrac{\bar{\sigma},\sigma}\\
	\delta\psi_\mu&=uD_\mu\sigma+v\sbrac{\phi_\mu,\sigma}\\
	\delta\widetilde{\psi}_\mu&=vD_\mu\sigma-u\sbrac{\phi_\mu,\sigma}.
	\end{aligned}
	\end{equation}
	These are the GL-twisted supersymmetry transformations which were found by Kapustin and Witten in \cite{kapustin2006electric}, and therefore 
	we have found a connection between our partially-twisted  5d \susy{2} SYM theory and the GL-twisted  4d \susy{4} SYM theory. In other words, \textit{we now have a 5d analogue of the GL-twist in 4d.}\footnote{Such a partial twist has also been discussed conceptually in \cite{Elliot}.}

	\section{Construction of The $\qcharge$-invariant Action}
	\subsection{$\cQ$-exact Action}
	
	We would like to find an action that is suitable for localization. 
	To do this, we need the action to be $\qcharge$-exact (where $\qcharge$ is some complex linear combination of the scalar supercharges we have considered) up to a metric-independent $\cQ$-invariant term which, as we shall see, turns out to take the form of the 4d Chern-Simons action. Before we write down the action, some observations are in order. 
	
	From \eqref{eqn:5dtwistedsusynew}, observe that the supersymmetry variation can be expressed as
	\begin{equation}
	\label{eqn:delta}
	\delta=u\delta_L+v\delta_R.
	\end{equation}
	Equivalently, we may write $\cQ$ in terms of supercharges $\qcharge_L$ and $\qcharge_R$, i.e.,
	\begin{equation}
	\qcharge=u\qcharge_L+v\qcharge_R,
	\end{equation}
	which acts on any field $\Phi$ as
	\begin{equation}
	\mbrac{\qcharge,\Phi}=\delta\Phi.
	\end{equation} 
	Here, we have added the labels $L$ and $R$ to remind ourselves of the corresponding left- and right-handed supersymmetries in the 4d case which were denoted by $\ell$ and $r$ in \cite{kapustin2006electric}. It should be emphasised that there is no concept of chirality in 5d. 
	In addition, we can rescale the supersymmetry variations such that they depend only on the ratio $t=v/u$, i.e., we obtain
	\begin{equation}
	\label{eqn:deltat}
	\delta_t=\delta_L+t\delta_R
	\end{equation}
	by dividing by $u$ on both sides of (\ref{eqn:delta}), and taking $\frac{\delta}{u}=\delta_t$. 
	
	Now, to put the action in $\qcharge$-exact form, we require that the transformation $\delta_t$ is nilpotent off-shell (up to gauge transformations). 
	In order to achieve this, we introduce some auxiliary fields. Let us first define two auxiliary $\mathfrak{g}$-valued 1-forms $H$ and $\widetilde{H}$, which modify the supersymmetry transformations of $\chi$ and $\widetilde{\chi}$ in (\ref{eqn:5dtwistedsusynew}). At the same time, the variations in $H$ and $\widetilde{H}$ may also be defined. Collectively we have, in terms of $\delta_t$,
	\begin{equation}
	\begin{aligned}
	\delta_t\chi_\alpha&=H_\alpha\\
	\delta_t\widetilde{\chi}_\alpha&=\widetilde{H}_\alpha\\
	\delta_t H_\alpha&=-i\cbrac{1+t^2}[\sigma,\chi_\alpha]\\
	\delta_t \widetilde{H}_\alpha&=-i\cbrac{1+t^2}[\sigma,\widetilde{\chi}_\alpha],
	\end{aligned}
	\end{equation}
	where $\alpha=1,2,3$ as before. We will construct an action whose equations of motion will impose the conditions $H=\mathcal{V}$ and $\widetilde{H}=t\widetilde{\mathcal{V}}$, where $\mathcal{V}$ and $\widetilde{\mathcal{V}}$ are functions of $t$ defined as 
	\begin{equation}\label{vs}
	\begin{aligned}
	&\mathcal{V}_\alpha(t)=\half \cbrac{\sbrac{F_{\alpha 4}
			+D_5\phi_\alpha
			+\half\varepsilon_{\alpha\beta\gamma}\cbrac{F^{\beta\gamma}-\sbrac{\phi^\beta,\phi^\gamma}}}+ t\sbrac{F_{\alpha 5}-D_4\phi_\alpha+\varepsilon_{\alpha\beta\gamma}D^\beta\phi^\gamma}}\\
	&\widetilde{\mathcal{V}}_\alpha(t)=\half \cbrac{\sbrac{F_{\alpha 4}+D_5\phi_\alpha-\half\varepsilon_{\alpha\beta\gamma}\cbrac{F^{\beta\gamma}-\sbrac{\phi^\beta,\phi^\gamma}}}- t^{-1}\sbrac{F_{\alpha 5}-D_4\phi_\alpha-\varepsilon_{\alpha\beta\gamma}D^\beta\phi^\gamma}},
	\end{aligned}
	\end{equation}
	and thereby there is on-shell agreement with the original supersymmetry transformations. 
	
	We may also define a $\mathfrak{g}$-valued $0$-form $P$, which modifies the supersymmetry transformations (\ref{eqn:5dtwistedsusynew}) of $\eta$ and $\widetilde{\eta}$. As with $H$ and $\widetilde{H}$, we may also define the corresponding supersymmetry transformation for $P$. Collectively we have
	\begin{equation}
	\begin{aligned}
	\delta\eta&=tP+\sbrac{\bar{\sigma},\sigma}\\
	\delta\widetilde{\eta}&=-P+t\sbrac{\bar{\sigma},\sigma}\\
	\delta P&=-it[\sigma,\eta]+i\sbrac{\sigma,\widetilde{\eta}}.
	\end{aligned}
	\end{equation}
	We shall require an action that imposes the equation of motion $P=F_{45}+D_\alpha\phi^\alpha$ to ensure that we have on-shell agreement with the original supersymmetry transformations. 

The introduction of the auxiliary fields gives rise to the following off-shell supersymmetry algebra for any field $\Phi$:
\begin{equation}
\delta_t^2\Phi=-i(1+t^2)\mathcal{L}_{\si}(\Phi),
\end{equation}
where $\mathcal{L}_{\si}(\Phi)$ is the change in $\Phi$ due to a gauge transformation generated by $\si$, to first order.  Furthermore, from \eqref{eqn:deltat}, we find that the algebra is equivalent to 
\begin{equation}\label{alg2}
\begin{aligned}
\delta_L^2\Phi=\delta_R^2\Phi&=-i\mathcal{L}_{\si}(\Phi),\\
\{\delta_L,\delta_R\}\Phi&=0.
\end{aligned}
\end{equation}	
	
	Now, for a gauge invariant fermionic expression, $\til{V}$, we can define a $\qcharge$-exact action to take the form $\{\qcharge,\widetilde{V}\}=\delta_t \widetilde{V}$. Our choice of $\til{V}$ has two parts, i.e., $\til{V}=\til{V}_1+\til{V}_2$.
	We shall first 
	pick
	\begin{equation}
	\widetilde{V}_1={2\over {g_5}^2}\int_{\cM} d^5x\ {4\over 1+t^2}\tr\cbrac{\chi_\alpha\cbrac{\half H^\alpha-\mathcal{V}^\alpha}+\widetilde{\chi}_{\al}\cbrac{\half \widetilde{H}^\alpha-t\widetilde{\mathcal{V}}^\alpha}}.
	\end{equation}
	The corresponding action takes the form
	\begin{equation}
	\label{eqn:S1}
	\begin{aligned}
	S_1&=\delta_t\widetilde{V}_1\\
	&={1\over {g_5}^2}\int_{\cM} d^5x\ \tr\cbrac{{4\over 1+t^2}\cbrac{\half H^\alpha H_\alpha-H_\alpha \mathcal{V}^\alpha}+{4\over 1+t^2}\cbrac{\half \widetilde{H}^\alpha \widetilde{H}_\alpha-t\widetilde{H}_\alpha \widetilde{\mathcal{V}}^\alpha}}+\ldots,
	\end{aligned}
	\end{equation}
	where the ellipsis indicates fermion terms that have been suppressed.
	Upon integrating out the auxiliary degrees of freedom, the ensuing action, including fermion terms, is
	\begin{equation}
	\label{eqn:S1'}
	\begin{aligned}
	S_1&={1\over {g_5}^2}\int_{\cM} d^5x\ \tr\left({-4\over 1+t^2}\cbrac{\mathcal{V}^\alpha\mathcal{V}_\alpha+t^2\widetilde{\mathcal{V}}^\alpha\widetilde{\mathcal{V}}_\alpha}+4i\chi_\alpha[\sigma,\chi^\alpha]+4i\widetilde{\chi}_\alpha[\sigma,\widetilde{\chi}^{\al}]\right.\\
	&\qquad\qquad\qquad\qquad\phantom{\half}+4\chi^\alpha \left[\phantom{\half}iD_\alpha\Upsilon-iD_4\psi_\alpha-iD_5\widetilde{\psi}_\alpha-i[\widetilde{\Upsilon},\phi_\alpha]\right.\\
	&\qquad\qquad\qquad\qquad\phantom{\half}\left.+{i\over 2}\varepsilon_{\alpha\beta\gamma}\cbrac{D^\beta\psi^\gamma +[\widetilde{\psi}^\beta,\phi^\gamma]-D^\gamma\psi^\beta-[\widetilde{\psi}^\gamma,\phi^\beta]}\right]\\
	&\qquad\qquad\qquad\qquad\phantom{\half} +4\widetilde{\chi}^\alpha \left[\phantom{\half}iD_\alpha\widetilde{\Upsilon}-iD_4\widetilde{\psi}_\alpha+iD_5\psi_\alpha+i[\Upsilon,\phi_\alpha]\right.\\
	&\qquad\qquad\qquad\qquad\left.\phantom{\half}\left.-{i\over 2}\varepsilon_{\alpha\beta\gamma}\cbrac{D^\beta\widetilde{\psi}^\gamma -[\psi^\beta,\phi^\gamma]-D^\gamma\widetilde{\psi}^\beta+[\psi^\gamma,\phi^\beta]}\right] \right).
	\end{aligned}
	\end{equation}

	Next, instead of directly constructing a $\qcharge$-exact action of the form $\delta_t \widetilde{V}_2$, we can make use of the fact that for a gauge invariant expression, $\til{V}'_2$, we have
	\begin{equation}
	\delta_L\delta_R \til{V}'_2= - \frac{1}{2t}(\delta_L+t\delta_R)(\delta_L-t\delta_R)\til{V}'_2,
	\end{equation}
	via \eqref{alg2}. 
	This means that we can also write down a $\qcharge$-exact action in the form $\delta_L\delta_R \til{V}'_2$. We shall do this to construct the second part of the $\cQ$-exact action, $\delta_t \widetilde{V}_2$. In other words, we have $\widetilde{V}_2=-\frac{1}{2t}(\delta_L-t\delta_R)\til{V}'_2$.
	We pick the gauge-invariant expression $\widetilde{V}_2'$ to be
	\begin{equation}
	\widetilde{V}_2'={2\over {g_5}^2}\int_{\cM} d^5x\ \tr\cbrac{-\half \eta\widetilde{\eta}-i\bar{\sigma}\cbrac{F_{45}+D_\alpha\phi^\alpha}},
	\end{equation}
	for which the corresponding action is
	\begin{equation}
	\label{eqn:S2}
	\begin{aligned}
	S_2&=\delta_L\delta_R \widetilde{V}_2\\
	&={2\over {g_5}^2}\int_{\cM} d^5x\ \tr\left(\half P^2-P\cbrac{F_{45}+D_\alpha\phi^\alpha}-D_M\bar{\sigma}D^M\sigma+\half[\bar{\sigma},\sigma]^2-[\phi_\alpha,\sigma][\phi^\alpha,\bar{\sigma}]\right.+\del_{\alpha}(\bar{\si}D^{\alpha}\si)\\
	&\qquad\qquad\qquad\qquad\phantom{\half}+i\widetilde{\eta}D_\alpha\widetilde{\psi}^\alpha+i\eta D_\alpha\psi^\alpha+ i\widetilde{\eta}\cbrac{D_4\widetilde{\Upsilon}+D_5\Upsilon}
	+i\eta\cbrac{D_4\Upsilon-D_5\widetilde{\Upsilon}}\\
	&\qquad\qquad\qquad\qquad\phantom{\half}-{i\over 2}[\sigma,\widetilde{\eta}]\widetilde{\eta}-{i\over 2}[\sigma,\eta]\eta-i\widetilde{\eta}[\psi_\alpha,\phi^\alpha]
	+i\eta[\widetilde{\psi}_\alpha,\phi^\alpha]+i[\bar{\sigma},\psi_\alpha]\psi^\alpha+i[\bar{\sigma},\widetilde{\psi}_\alpha]\widetilde{\psi}^\alpha\\
	&\qquad\qquad\qquad\qquad\left.\phantom{\half}+i[\bar{\sigma},\Upsilon]\Upsilon+i[\bar{\sigma},\widetilde{\Upsilon}]\widetilde{\Upsilon}\right).
	\end{aligned}
	\end{equation}
	Here, we can integrate $P$ out to obtain
	\begin{equation}
	\label{eqn:S2'}
	\begin{aligned}
	S_2&={2\over {g_5}^2}\int_\cM d^5x\ \tr\left(-\half \cbrac{F_{45}+D_\alpha\phi^\alpha}^2-D_M\bar{\sigma}D^M\sigma+\half[\bar{\sigma},\sigma]^2-[\phi_\alpha,\sigma][\phi^\alpha,\bar{\sigma}]\right.+\del_{\alpha}(\bar{\si}D^{\alpha}\si)\\
	&\qquad\qquad\qquad\qquad\phantom{\half}+i\widetilde{\eta}D_\alpha\widetilde{\psi}^\alpha+i\eta D_\alpha\psi^\alpha+ i\widetilde{\eta}\cbrac{D_4\widetilde{\Upsilon}+D_5\Upsilon}
	+i\eta\cbrac{D_4\Upsilon-D_5\widetilde{\Upsilon}}\\
	&\qquad\qquad\qquad\qquad\phantom{\half}-{i\over 2}[\sigma,\widetilde{\eta}]\widetilde{\eta}-{i\over 2}[\sigma,\eta]\eta-i\widetilde{\eta}[\psi_\alpha,\phi^\alpha]
	+i\eta[\widetilde{\psi}_\alpha,\phi^\alpha]+i[\bar{\sigma},\psi_\alpha]\psi^\alpha+i[\bar{\sigma},\widetilde{\psi}_\alpha]\widetilde{\psi}^\alpha\\
	&\qquad\qquad\qquad\qquad\left.\phantom{\half}+i[\bar{\sigma},\Upsilon]\Upsilon+i[\bar{\sigma},\widetilde{\Upsilon}]\widetilde{\Upsilon}\right).
	\end{aligned}
	\end{equation}
	
	The full $\qcharge$-exact action is just the sum of the actions in (\ref{eqn:S1'}) and (\ref{eqn:S2'}), and can be written as
	\begin{equation}
	\label{eqn:S}
	\begin{aligned}
	S&=S_1+S_2\\
	&={1\over {g_5}^2}\int_\cM d^5x\ \tr\left({-4\over 1+t^2}\cbrac{\mathcal{V}^\alpha\mathcal{V}_\alpha+t^2\widetilde{\mathcal{V}}^\alpha\widetilde{\mathcal{V}}_\alpha}+4i\chi_\alpha[\sigma,\chi^\alpha]+4i\widetilde{\chi}_\alpha[\sigma,\widetilde{\chi}]\right.\\
	&\qquad\qquad\qquad\qquad\phantom{\half}+4\chi^\alpha \left[\phantom{\half}iD_\alpha\Upsilon-iD_4\psi_\alpha-iD_5\widetilde{\psi}_\alpha-i[\widetilde{\Upsilon},\phi_\alpha]\right.\\
	&\qquad\qquad\qquad\qquad\phantom{\half}\left.+{i\over 2}\varepsilon_{\alpha\beta\gamma}\cbrac{D^\beta\psi^\gamma +[\widetilde{\psi}^\beta,\phi^\gamma]-D^\gamma\psi^\beta-[\widetilde{\psi}^\gamma,\phi^\beta]}\right]\\
	&\qquad\qquad\qquad\qquad\phantom{\half} +4\widetilde{\chi}^\alpha \left[\phantom{\half}iD_\alpha\widetilde{\Upsilon}-iD_4\widetilde{\psi}_\alpha+iD_5\psi_\alpha+i[\Upsilon,\phi_\alpha]\right.\\
	&\qquad\qquad\qquad\qquad\phantom{\half}\left.-{i\over 2}\varepsilon_{\alpha\beta\gamma}\cbrac{D^\beta\widetilde{\psi}^\gamma -[\psi^\beta,\phi^\gamma]-D^\gamma\widetilde{\psi}^\beta+[\psi^\gamma,\phi^\beta]}\right]\\
	&\qquad\qquad\qquad\qquad\phantom{\half}- \cbrac{F_{45}+D_\alpha\phi^\alpha}^2-2D_M\bar{\sigma}D^M\sigma+[\bar{\sigma},\sigma][\bar{\sigma},\sigma]-2[\phi_\alpha,\sigma][\phi^\alpha,\bar{\sigma}]+2\del_{\alpha}(\bar{\si}D^{\alpha}\si)\\
	&\qquad\qquad\qquad\qquad\phantom{\half}+2i\widetilde{\eta}D_\alpha\widetilde{\psi}^\alpha+2i\eta D_\alpha\psi^\alpha+ 2i\widetilde{\eta}\cbrac{D_4\widetilde{\Upsilon}+D_5\Upsilon}
	+2i\eta\cbrac{D_4\Upsilon-D_5\widetilde{\Upsilon}}\\
	&\qquad\qquad\qquad\qquad\phantom{\half}-{i}[\sigma,\widetilde{\eta}]\widetilde{\eta}-{i}[\sigma,\eta]\eta-2i\widetilde{\eta}[\psi_\alpha,\phi^\alpha]
	+2i\eta[\widetilde{\psi}_\alpha,\phi^\alpha]+2i[\bar{\sigma},\psi_\alpha]\psi^\alpha+2i[\bar{\sigma},\widetilde{\psi}_\alpha]\widetilde{\psi}^\alpha\\
	&\qquad\qquad\qquad\qquad\left.\phantom{\half}+2i[\bar{\sigma},\Upsilon]\Upsilon+2i[\bar{\sigma},\widetilde{\Upsilon}]\widetilde{\Upsilon}\right).
	\end{aligned}
	\end{equation}
	Next, we make use of the identity
	\begin{equation}
	\label{id:t-dependence}
	\begin{aligned}
	&\phantom{=}{1\over {g_5}^2}\int_\cM d^5x\ \tr\left({-4t^{-1}\over t+t^{-1}}\cbrac{\mathcal{V}^\alpha\mathcal{V}_\alpha+t^2\widetilde{\mathcal{V}}^\alpha\widetilde{\mathcal{V}}_\alpha}-\cbrac{F_{45}+D_\alpha\phi^\alpha}^2\right)\\
	&=-{1\over {g_5}^2}\int_\cM d^5x\ \tr\left(F_{\alpha m}F^{\alpha m}+F_{45}F^{45}+\half F_{\alpha\beta}F^{\alpha\beta}+D_m\phi_\alpha D^m\phi^\alpha+D_\alpha\phi_\beta D^\alpha\phi^\beta\right.\\
	&\qquad\qquad\qquad\qquad\phantom{\half}+\half[\phi_\al,\phi_\bt][\phi^\al,\phi^\bt]+\partial_\alpha\cbrac{\phi^\alpha D_\beta\phi^\beta}-\partial_\gamma\cbrac{\phi_\delta D^\delta\phi^\gamma}+2\del_{\al}(F_{45}\phi^\al)\\
	&\qquad\qquad\qquad\qquad\phantom{\half} -4\cbrac{{t-t^{-1}\over t+t^{-1}}}\cbrac{\half\varepsilon^{\alpha\beta\gamma}}\cbrac{\half F_{\alpha 4}F_{\beta\gamma}+\half\partial_\alpha\cbrac{\phi_\beta D_4\phi_\gamma}+\partial_\alpha\cbrac{F_{\beta 5}\phi_\gamma}}\\
	&\qquad\qquad\qquad\qquad\left.\phantom{\half}+\cbrac{{8\over t+t^{-1}}}\cbrac{\half\varepsilon^{\alpha\beta\gamma}}\cbrac{\half F_{\alpha 5}F_{\beta\gamma}+\half\partial_\alpha\cbrac{\phi_\beta D_5\phi_\gamma}-\partial_\alpha\cbrac{F_{\beta 4}\phi_\gamma}}\right),
	\end{aligned}
	\end{equation}
	to rewrite the action (\ref{eqn:S}) in terms of a $t$-independent part and a $t$-dependent part. Note that, apart from total derivative terms and $t$-dependent terms, the boson terms in the action correspond to standard kinetic and potential terms of 5d $\mathcal{N}=2$ supersymmetric Yang-Mills, partially-twisted along $Y\times \R_+$.
	
	Now, the $t$-dependent term of the $\cQ$-exact action takes the form
	\begin{equation}
	\label{eqn:St}
	\begin{aligned}
	S_t&={1\over {g_5}^2}\int_\cM d^5x\ \varepsilon^{\alpha\beta\gamma}\tr\left(2\cbrac{{t-t^{-1}\over t+t^{-1}}}\cbrac{\half F_{\alpha 4}F_{\beta\gamma}+\half\partial_\alpha\cbrac{\phi_\beta D_4\phi_\gamma}+\partial_\alpha\cbrac{F_{\beta 5}\phi_\gamma}}\right.\\
	&\qquad\qquad\qquad\qquad\qquad\left.\phantom{\half}-\cbrac{{4\over t+t^{-1}}}\cbrac{\half F_{\alpha 5}F_{\beta\gamma}+\half\partial_\alpha\cbrac{\phi_\beta D_5\phi_\gamma}-\partial_\alpha\cbrac{F_{\beta 4}\phi_\gamma}}\right).
	\end{aligned}
	\end{equation}
Since we do not expect such a $t$-dependent term from the physical 5d action, we shall choose to cancel it by adding 
\begin{equation}\label{s2}
S_3=-S_t
\end{equation}
to the action. Therefore, our total action is now $S_1+S_2+S_3$. However, this action is not yet $\cQ$-invariant, which leads us to discuss boundary conditions for this action that will ensure $\cQ$-invariance.
	
	
	\subsection{Boundary Conditions from NS5-branes}
	

We shall now specify boundary conditions at the origin of $\R_+$, i.e., $x^3=0$, such that we have a system that can be understood as the worldvolume theory of a stack of D4-branes ending on a (deformed) NS5-brane in type IIA string theory. Specifically, we consider the following configuration in flat Euclidean space :

\vspace{3em}
\begin{tikzpicture}[overlay]
\draw [decorate,decoration={brace,amplitude=4pt},xshift=61pt,yshift=0pt]
(2.65,-0.2) -- (3.79,-0.2) node [black,midway,yshift=9pt] 
{\footnotesize $Y$};
\draw [decorate,decoration={brace,amplitude=4pt},xshift=61pt,yshift=0pt]
(4.05,-0.2) -- (4.57,-0.2) node [black,midway,yshift=9pt] 
{\footnotesize $\mathbb{R}$};
\draw [decorate,decoration={brace,amplitude=4pt},xshift=61pt,yshift=0pt]
(4.83,-0.2) -- (5.97,-0.2) node [black,midway,yshift=9pt] 
{\footnotesize $\Sigma$};
\draw [decorate,decoration={brace,amplitude=4pt},xshift=61pt,yshift=0pt]
(6.23,-0.2) -- (8.04,-0.2) node [black,midway,yshift=9pt] 
{\footnotesize $N\tilde{V}\!\!\subset \!\!T^\ast \tilde{V}$};
\draw [decorate,decoration={brace,amplitude=4pt},xshift=61pt,yshift=0pt]
(2.65,0.4) -- (4.57,0.4) node [black,midway,yshift=9pt] 
{\footnotesize $\tilde{V}$};
\end{tikzpicture}
\begin{center}
\begin{tabular}{@{}lllllllllll@{}}
\toprule
   & \textbf{1} & \textbf{2} & \textbf{3} & \textbf{4} & \textbf{5} & \textbf{6} & \textbf{7} & \textbf{8} & \textbf{9} & \textbf{10}\\
   \textbf{D4} & \x & \x & \x & \x & \x & & & & &\\
   $\til{\textrm{\textbf{NS5}}}$ & \x & \x &  & \x & \x & \x & \x & & & \\
\bottomrule
 \end{tabular}
\end{center}
where, e.g., an empty entry under `3' indicates that the brane is located at $x^3=0$. The scalar fields $\{\phi_{\wht{1}},\phi_{\wht{2}},\phi_{\wht{3}},\phi_{\wht{4}},\phi_{\wht{5}}\}$ of the 5d theory are understood to parametrize the $\{6,7,8,9,10\}$ directions, respectively. Note that the partial twist arises in this configuration because $V\subset\tilde{V}=Y\times \R$, where $\tilde{V}$ is the zero section of the cotangent bundle $T^*\tilde{V}$, and `coordinates' normal to $\tilde{V}$ in $T^*\tilde{V}$ must be components of one-forms, as we obtained via twisting \cite{bersh}.

	
	Firstly, this configuration implies that the fields $\phi_3$, $\sigma$ and $\ov{\si}$ obey Dirichlet boundary conditions which set them to zero at the boundary.
	Secondly, the sets of fields \{$\phi_1$,$\phi_2$\} and \{$A_1$,$A_2$,$A_4$,$A_5$\} obey generalized Neumann boundary conditions 
	that can be obtained from boundary interactions given by
	\begin{equation}\label{liftedboundaryint}
	\begin{aligned}
	S_{\partial \mathcal{M}}&={1\over {g_5}^2}\int_{\partial \mathcal{M}} d^4x\ \tr\left(\cbrac{t+t^{-1}}\cbrac{\half\varepsilon^{{\tal}{\tbt} }D_5\phi_{\tal}\phi_{\tbt} }\right.\\
	&\qquad\qquad\qquad\qquad\left.\phantom{\half}+\cbrac{{t+t^{-1}\over t-t^{-1}}}\varepsilon^{ijk}\cbrac{A_i\partial_j A_k+{2\over 3}A_i A_j A_k}\right),
	\end{aligned}
	\end{equation}
by insisting that the equations of motion do not have boundary corrections. 
Here, the indices $i,j,k=1,2,4$ and $\tal,\tbt=1,2$, and the Levi-Civita symbols are defined such that $\varepsilon^{124}=1$ and $\varepsilon^{12}=1$.
	
	Thirdly, the lifted boundary conditions on the fermionic fields are projection conditions that follow from supersymmetry.
We also note that the boundary conditions above imply that 
	\begin{equation}
	\begin{aligned}
	\delta(A_i+w\phi_i)&=0\\
	\delta(A_4+wA_5)&=0,
	\end{aligned}
	\end{equation}
	for $i=1,2$ and $w={t-t^{-1}\over 2}$, along the boundary. Finally, the boundary conditions restrict the complex parameter $t$ such that $|t|=1$.
	
Dimensional reduction, as detailed in Section 2.3, of the boundary conditions above will indeed reduce them to the GL-twist of the (deformed) NS5 boundary conditions for 4d $\cN=4$ supersymmetric Yang-Mills theory	
	derived in \cite{GW} and used in \cite{witten2011fivebranes}. 
Note that the constraint $|t|=1$ is satisfied in the 4d theory since it obeys $t^2=\frac{\ov{\tau}}{\tau}$, where $\tau$ is the complex coupling parameter of this theory. 

\subsection{5d Topological-holomorphic Theory}
	
	Having specified the boundary conditions and boundary interactions at $x_3=0$ for our partially-twisted theory, we first note that the total derivative terms $\frac{2}{g_5^2}\int_\cM d^5x~\textrm{Tr}~\del_{\alpha}(\bar{\si}D^{\alpha}\si)$ in \eqref{eqn:S} and $-\frac{1}{g_5^2}\int_\cM d^5x~\textrm{Tr}~\big(\partial_\alpha\cbrac{\phi^\alpha D_\beta\phi^\beta}-\partial_\gamma\cbrac{\phi_\delta D^\delta\phi^\gamma}+2\del_{\al}(F_{45}\phi^\al)\big)$ in \eqref{id:t-dependence} vanish via Stoke's theorem and the boundary conditions $\bar{\si}=0$ and $\phi_3=0$. It can be checked that the action we have dimensionally reduces exactly to the GL-twisted 4d $\mathcal{N}=4$ super Yang-Mills action studied by Witten in \cite{witten2011fivebranes}, via the procedure given in Section 3.
	
	We would like to further understand our partially-twisted theory, in particular the non-$\cQ$-exact sector. Recall that along the boundary, we have $t$-dependent boundary interactions given by \eqref{liftedboundaryint}.  
	Miraculously, when these boundary interactions are combined with the $t$-dependent action $S_3$ in (\ref{s2}), a term proportional to a 4d Chern-Simons action is obtained. To see this, we first need to rewrite the coordinates ${x^4,x^5}\in\Sigma$ in terms of complex ones. We define the coordinates
	\begin{equation}\label{comp}
	\begin{aligned}
	&z_w=2\frac{(wx^4-x^5)}{w-\ov{w}}\\
	&\bar{z}_w=2\frac{(\ov{w}x^4-x^5)}{\ov{w}-w},
	\end{aligned}
	\end{equation}
	where $w$ was defined previously to be $w=\frac{t-t^{-1}}{2}$. These coordinates are chosen since we shall make use of one of their corresponding partial derivatives, which are
	\begin{equation}\label{comp}
	\begin{aligned}
	&\del_{z_w}=\frac{1}{2}(\del_4+\ov{w}\del_5)\\
	&\del_{\ov{z}_w}=\frac{1}{2}(\del_4+{w}\del_5).
	\end{aligned}
	\end{equation}
	 We also introduce the complexified gauge fields 
	\begin{equation}
	\cA_{w \tal}=A_{\tal}+w\phi_{\tal}
	\end{equation}
	(for $\tal=1,2$) and
	\begin{equation}
	\cA_{w\zb_w}=\half\cbrac{A_4+wA_5}
	\end{equation}
that are $\cQ$-invariant along the boundary.
 	We may then write the combined $t$-dependent action as
 	\begin{equation}\label{long}
 	\begin{aligned}
 		S_3+S_{\partial M}={i\Psit\over 4\pi}\int_{\del \cM} d^4x~\textrm{Tr }\big(&\cA_{w1}(\del_2A_4-\del_4\cA_{w2})+\frac{2}{3}\cA_{w1}[\cA_{w2},A_4]\\
 	&+\cA_{w2}(\del_4\cA_{w1}-\del_1 A_{4})+\frac{2}{3}\cA_{w2}[A_{4},\cA_{w1}]\\
 	&+A_{4}(\del_1\cA_{w2}-\del_2 \cA_{w1})+\frac{2}{3}A_{4}[\cA_{w1},\cA_{w2}]
 	\big)\\
 	+w{i\Psit\over 4\pi}\int_{\del \cM} d^4x~&\textrm{Tr }\big(\cA_{w1}(\del_2A_5-\del_5\cA_{w2})+\frac{2}{3}\cA_{w1}[\cA_{w2},A_5]\\
 	&+\cA_{w2}(\del_5\cA_{w1}-\del_1 A_{5})+\frac{2}{3}\cA_{w2}[A_{5},\cA_{w1}]\\
 	&+A_{5}(\del_1\cA_{w2}-\del_2 \cA_{w1})+\frac{2}{3}A_{5}[\cA_{w1},\cA_{w2}]
 	\big),
 	\end{aligned}
 	\end{equation}
(details of the derivation can be found in the appendix), or more succintly as	
	\begin{equation}\label{4dcsw}
	S_3+S_{\partial M}={w-\bar{w} \over 4}{i\Psit\over 2\pi}\int_{\partial M}\ dz_w\wedge\tr\cbrac{\cA_w\wedge d\cA_w+{2\over 3}\cA_w\wedge \cA_w\wedge \cA_w},
	\end{equation}
	where $\Psit$ is a $t$-dependent parameter, which may be written as
	\begin{equation}
	\Psit={4\pi i\over {g_5}^2}\cbrac{{t-t^{-1}\over t+t^{-1}}-{t+t^{-1}\over t-t^{-1}}}.
	\end{equation}
	
	Hence, we have found an action that is $\cQ$-exact up to the $\cQ$-invariant term \eqref{4dcsw}, which is explicitly given by

\begin{equation}\label{topholo}
\boxed{
S=\delta_t\til{V}_1-\frac{1}{2t}\delta_t(\delta_L-t\delta_R)\til{V}'_2+{w-\bar{w} \over 4}{i\Psit\over 2\pi}\int_{\partial \cM}\ dz_w\wedge\tr\cbrac{\cA_w\wedge d\cA_w+{2\over 3}\cA_w\wedge \cA_w\wedge \cA_w}.
}\end{equation}	
	This 5d theory is \textit{topological-holomorphic}, with the boundary action depending on the complex structure defined on $\Sigma$ via \eqref{comp}.
	
	\section{Localization To 4d Chern-Simons Theory and Integrable Lattice Models}
	We shall now explain how the path integral of our 5d topological-holomorphic theory is equivalent to a path integral of 4d Chern-Simons theory that is valid beyond perturbation theory.
	                                           
  Firstly, in what follows, we shall exclude $t=\pm 1$ in the classical theory, as we require the parameter $w=\frac{t-t^{-1}}{2}$ to be nonzero to eventually obtain 4d Chern-Simons theory. In fact, since $|t|=1$, $w=\frac{t-t^{-1}}{2}$ is purely imaginary, i.e., $w=i\textrm{Im}(w)$. In addition, we shall also only consider $t\neq \pm i$. As a result, one can show that the $t$-dependence of $\delta_t$ can be eliminated via rescaling of $\delta_t$ as well as fermion redefinitions. 
  Consequently, as $t$ only appears in a $\cQ$-exact term, it is irrelevant for our topological-holomorphic theory. Also note that since $|t|=1$, $\Psit$ is real.
  
	Secondly, 
	 the path integral localizes to field configurations that obey $\delta_t \lambda=0$, for a fermionic field, $\lambda$. However, not all of these configurations play an equal role, as we shall see shortly. Via some field redefinitions, we can find fermionic fields whose (on-shell) variations are $\cV_{\alpha}(t)$, $\til{\cV}_{\alpha}(t)$ and $\cV_0=F_{45}+D_{\alpha}\phi^{\alpha}$, which we refer to as $\chi_{\al}$, $\til{\chi}'_{\al}$ and $\eta'$, respectively. Localization of the path integral to $\cV_{\al}=\til{\cV}_{\al}=\cV_0=0$ can be achieved 
by scaling up the $\cQ$-exact terms in \eqref{topholo}, since these equations are among the conditions for the $\cQ$-exact terms to vanish. Hence, we find that the path integral is supported on the solution space of the equations 
	\begin{equation}\label{bpsvs}
	\begin{aligned}
	\cV_{\alpha}(t)&=0\\ \til{\cV}_{\alpha}(t)&=0\\ \cV_0&=0.
\end{aligned}	
	\end{equation}
	 The remaining (bosonic) field configurations necessary for the $\cQ$-exact terms terms to vanish 
	 are 
	 \begin{equation}
	 \begin{aligned}
	 D_M\si&=0\\
	 [\phi_{\al},\si]&=0\\
	 [\si,\ov{\si}]&=0,
	 \end{aligned}
	 \end{equation}
 (since $t\neq \pm i$)	and these just imply that $\si=0$ everywhere on $\cM$, since we have imposed $\si=0$ at the boundary. 

	Now, localization of the path integral of the 5d partially-twisted theory reduces it to a path integral over the bosonic fields $\cA_{w\tal}$ and $\cA_{w\zb}$ along the boundary, with the integral being restricted to 
	solutions of the equations \eqref{bpsvs}. This follows since the bulk modes contained in the $\cQ$-exact action can be integrated out to give boson and fermion one-loop determinants that cancel due to the $\cQ$-symmetry, leaving only the path integral over the boundary action (assuming a certain anomaly vanishes, as we discuss later in this section).  
However, note that to see that the resulting path integral actually converges, we require that the parameter $t$ takes a suitable value in the $\cQ$-exact 
sector of the action prior to localization, via the addition of $\cQ$-exact terms. 
As we shall see, $t=\pm 1$ are such suitable values. Note that although we exclude these values classically, the freedom to add $\cQ$-exact terms to the action in the quantum theory allows $t=\pm 1$ in the resulting $\cQ$-exact sector.
	
	Another way to understand this localization is that our partially-twisted theory can be interpreted as a 1d gauged A-model, studied in \cite{anewlook}, with target space being the space $\mathfrak{A}$ of all possible $\cA_{w\tal}$ and $\cA_{w\zb}$ fields, and gauge group the space $H$ of maps  from $Y\times \Sigma$ to the worldvolume gauge group, $G$ (assuming that we formulate our 5d theory using a trivial $G$-bundle). The essential observation is that a 4d Chern-Simons action serves as the superpotential of the 1d theory, via which 5d bulk terms involving fields in $\mathfrak{A}$ can be obtained as standard terms of the 1d theory.
	
To see this explicitly, let us pick on $\mathfrak{A}$ the metric
	\begin{equation}\label{fieldmetric}
	g=-\frac{1}{2g_5^2}\int_{Y\times \Sigma}d^2z d^2x ~\textrm{Tr}(\delta\cA_{\til{\alpha}}\otimes \delta \ov{\cA}^{\til{\alpha}}+\delta\ov{\cA}_{\til{\alpha}}\otimes \delta {\cA}^{\til{\alpha}}+4\delta \cA_{\zb}\otimes \delta \cA_z+4\delta \cA_{z}\otimes \delta \cA_\zb),
	\end{equation} 
and the complex structure where $\cA_{\tal}$ and $\cA_{\zb}$ are holomorphic, which implies the following moment map for the $H$-action:
\begin{equation}
\mu=-\frac{1}{g_5^2}(D_{\til{\al}}\phi^{\til{\al}}+F_{45}).
\end{equation}
Here, we have defined the complex gauge fields \begin{equation}
\begin{aligned}
\cA_{\al}=A_{\al}+i\phi_{\al}, && \cAb_{\al}=A_{\al}-i\phi_{\al},
\end{aligned}
\end{equation}
and
\begin{equation}
\begin{aligned}
\cA_{z}=\frac{1}{2}(A_{4}-iA_5), && \cA_{\zb}=\frac{1}{2}(A_{4}+iA_5).
\end{aligned}
\end{equation}
Let us also define the 4d Chern-Simons superpotential  
\begin{equation}\label{1w}
W=-\frac{e^{i\al}}{g_5^2}\int_{Y\times \Sigma} dz\wedge \textrm{Tr}\bigg(\cA \wedge d\cA +{2\over 3}\cA \wedge \cA\wedge \cA\bigg).
\end{equation}
As we shall see, $W$ is not uniquely determined; as only derivatives of $W$ enter the 1d gauged A-model action, it is defined modulo an additive constant. Furthermore, it is only defined up to an arbitrary phase factor (that can be changed by an R-symmetry rotation), which we have denoted by $e^{i\al}$. 

Now, the 1d gauged A-model action has the form (suppressing fermion terms and boundary action for brevity) 
\begin{equation}
\begin{aligned}
S^{Bose}_{1d}=\int d\tau \big(& g_{i\ov{\jmath}}\del_{\tau}^A x^i\del_{\tau}^A x^{\ovj}+g_{i\ov{\jmath}}
V_a^i\tsi^{a}\ov{V}_b^{\ovj}\ov{\tsi}^b+g_{i\ov{\jmath}}
V_a^i\ov{\tsi}^{a}\ov{V}_b^{\ovj}{\tsi}^b+g_{i\ov{\jmath}}
V_a^i\tphi^{a}\ov{V}_b^{\ovj}{\tphi}^b\\&-g_{i\ov{\jmath}}F^i\ov{F}^{\ovj}+\frac{1}{2}F^i\del_iW+\frac{1}{2}\ov{F}^{\ovj}\del_{\ovj}\ov{W}\big)
\\ -\frac{1}{e^2}\int &d\tau~\textrm{Tr'}\big(D_{\tau}\tphi D_{\tau}\tphi+2D_{\tau}\tsi D_{\tau}\ov{\tsi}+[\tsi,\ov{\tsi}][\ov{\tsi},\tsi]+2[\tphi,\tsi][\tphi,\ov{\tsi}]\\&-D^2+2e^2	\mu D\big),
\end{aligned}
\end{equation}
where $\del_{\tau}^A x^i=\del_{\tau} x^i+A^a_{\tau}V_a^i$. Here, $x$ is a map from $\R_+$ to $\mathfrak{A}$, $V_a$, $a=1,\ldots, \textrm{dim }H$ are the Killing vector fields generating the action of $H$ on $\mathfrak{A}$, $\tphi^a$ is a real scalar field, $\til{\si}^a$ and $\ov{\til{\si}}^a$ are complex scalar fields, $F^i$ and $D$ are auxiliary fields, $e^2$ is a coupling constant, and $\textrm{Tr'}$ is a negative-definite quadratic form on the Lie algebra of $H$.

By integrating out the auxiliary fields $F^i$ and $D$, we 
find that the potential energy terms of the form $\int d{\tau}\bigg( \frac{1}{4}g^{i\ov{\jmath}}\del_iW\del_{\ov{\jmath}}\ov{W}-e^4\textrm{Tr'}\mu^2\bigg)$ of the 1d gauged sigma model are (for ${\tau}=x^3$)
\begin{equation}\label{btfm}
-\frac{1}{g_5^2}\int_{\cM} d^5x~\textrm{Tr}~\bigg(\frac{1}{2}\cF^{\tal\tbt}\cFb_{\tal\tbt}+4\cF^{\tal}_{~\zb}\cFb_{\tal z}+(-2i\cF_{z\zb}+D_{\tal}\phi^{\tal})^2\bigg),
\end{equation}	
where the covariant derivatives
\begin{equation}
\begin{aligned}
\cD_{\al}=\del_{\al}+[\cA_{\al},\cdot\textrm{ }], && \cDb_{\al}=\del_{\al}+[\cAb_{\al},\cdot\textrm{ }],
\end{aligned}
\end{equation}
and 
\begin{equation}
\begin{aligned}
\cD_{z}=\del_{z}+[\cA_{z},\cdot\textrm{ }], && \cD_{\zb}=\del_{\zb}+[\cA_{\zb},\cdot\textrm{ }],
\end{aligned}
\end{equation}
have been used to define the field strengths $\cF_{\bt\gm}=[\cD_{\bt},\cD_{\gm}]$, $\cF_{\al \zb}=[\cD_{\al},\cD_{\zb}]$ and $\cF_{z \zb}=[\cD_{z},\cD_{\zb}]$.
 Upon integration by parts, \eqref{btfm} is equal to 
\begin{equation}\label{uo}
-\frac{1}{g_5^2}\int_{\cM} d^5x~\textrm{Tr}~\bigg(\frac{1}{2}F^{\tal\tbt}F_{\tal\tbt}+D^{\tal}\phi^{\tbt}D_{\tal}\phi_{\tbt}+\frac{1}{2}[\phi^{\tal},\phi^{\tbt}][\phi_{\tal},\phi_{\tbt}]+4F^{\tal}_{~~\zb}F_{\tal z}+4D_z\phi_{\tal}D_{\zb}\phi^{\tal}-4\cF_{z\zb}\cF_{z\zb}\bigg).
\end{equation}		
These are just the bulk boson terms in our partially-twisted 5d action that do not involve the $x^3$ direction nor the fields, $\phi_3$, $\si$ and $\bar{\si}$. 
The remaining bulk boson terms of the 1d theory correspond to the 5d bulk boson terms not given in \eqref{uo}, i.e., by identifying 
$e^2$, $\tphi$, $\tsi$ and $\ov{\tsi}$ with $g_5^2$, $\phi_3$, $\si$ and $\ov{\si}$ respectively, the remaining terms are
\begin{equation}
\begin{aligned}
-\frac{1}{g_5^2}\int_{\cM} d^5x~\textrm{Tr}~\bigg(&F_{3\tal}F^{3\tal}+D_{3}\phi_{\tal}D^3\phi^{\tal}+F_{3x}F^{3x}+2D_{\tal}\si D^{\tal}\ov{\si}+2[\phi_{\tal},\si][\phi^{\tal},\ov{\si}]+2D_{\til{z}}\si D^{\til{z}}\ov{\si}\\
&+D_{\tal}\phi_3D^{\tal}\phi^3+[\phi_{\tal},\phi_3][\phi^{\tal},\phi^3]+D_{\til{z}}\phi_3D^{\til{z}}\phi^3\\
&+D_3\phi_3D^3\phi^3+2D_3\si D^3\ov{\si}+[\si,\ov{\si}][\ov{\si},\si]+2[\phi_3,\si][\phi^3,\ov{\si}]\bigg),
\end{aligned}
\end{equation}
where the index ${\til{z}}={z,\zb}$. In this manner, the entire 5d topological-holomorphic theory can be shown to be equivalent to the 1d gauged A-model with target $\mathfrak{A}$, and boundary action \eqref{4dcsw}.

Hence, from \cite{anewlook}, we know that this model should localize to the boundary action. However, note that there is an anomaly arising from the residual $SO(2)$ R-symmetry that does not enter the twisting. We shall assume that this anomaly vanishes; even when it does not, we may insert suitable operators to guarantee a non-vanishing path integral. We also have to specify the boundary conditions at $x^3=\infty$ on the half-line, which we take to be a critical point of the 4d Chern-Simons action that satisfies $\mu=0$, together with Neumann boundary conditions on $\phi_3$, $\si$ and $\ov{\si}$ (note that $\mu=0$ and the Neumann boundary condition for $\si$ also sets  to zero the total derivative terms discussed at the beginning of Section 3.3). This ensures that the contribution of the boundary theory at $x^3=\infty$ is just an overall constant in the path integral, which can be absorbed into the measure.

 We thus arrive at 
\begin{equation}\label{exppp}
\int_{\til{\Gamma}} D\cA_w~\textrm{exp}\Bigg(-{w-\bar{w} \over 4}{i\Psit\over 2\pi}\int_{\partial \cM}\ dz_w\wedge\tr\cbrac{\cA_w\wedge d\cA_w+{2\over 3}\cA_w\wedge \cA_w\wedge \cA_w}\Bigg),
\end{equation}
where $\til{\Gamma}$ is a subspace of $\mathfrak{A}$ 
defined by solutions of 
 \eqref{bpsvs}.
Recalling that the argument of the exponent in 
\eqref{exppp} is the negative of \eqref{long}, we shall perform the coordinate redefinition
\begin{equation}
x^5\rightarrow\textrm{Im}(w){x}^5,
\end{equation}
that implies $d^4x\rightarrow \textrm{Im}(w) d^4x$, $\textrm{Im}(w) \del_5\rightarrow{\del}_5$ and $\textrm{Im}(w) A_5\rightarrow{A}_5$, whereby this argument becomes 
\begin{equation}
\begin{aligned}
-\frac{i\Psit}{2\pi}~\textrm{Im}(w)\int_{\del\cM}d^4x~\textrm{Tr}(&\cA_{w1}(\del_2\cA_{\zb}-\del_{\zb}\cA_{w2})+\frac{2}{3}\cA_{w1}[\cA_{w2},A_{\zb}]\\
 	&+\cA_{w2}(\del_{\zb}\cA_{w1}-\del_1 A_{\zb})+\frac{2}{3}\cA_{w2}[A_{\zb},\cA_{w1}]\\
 	&+A_{\zb}(\del_1\cA_{w2}-\del_2 \cA_{w1})+\frac{2}{3}A_{\zb}[\cA_{w1},\cA_{w2}]).
\end{aligned}
\end{equation}
Now, apart from a factor multiplying the action, and a factor multiplying the path integral measure (that we can remove via a choice of normalization), the path integral only depends on $\textrm{Im}(w)$ in the definitions of $\cA_{w1}$ and $\cA_{w2}$. Hence, we can conveniently 
fix it in these fields, whereby we obtain the path integral 
\begin{equation}\label{4dcsf}
\boxed{
\int_{\til{\Gamma}} D\cA~\textrm{exp}\Bigg({\Psit \textrm{Im}(w)\over 4\pi}\int_{\partial \cM}\ dz\wedge\tr\cbrac{\cA\wedge d\cA+{2\over 3}\cA\wedge \cA\wedge \cA}\Bigg).}
\end{equation}	
 \textit{This is the path integral for 4d Chern-Simons theory with} $\frac{1}{\hbar}=-\frac{i\Psit \textrm{Im}(w)}{2}$\textit{, defined beyond perturbation theory with an integration cycle determined by $\til{\Gamma}$.
	} 
	 This integration cycle in fact ensures the convergence of the path integral, 
	 as long as we tune the value of $t$ in the $\cQ$-exact terms of the action (prior to localization) to $1$ or $-1$.
To see this, note that the equations  $\delta\chi_{\al}=\cV_{\al}=0$ and $\delta\til{\chi}'_{\al}=\til{\cV}_{\al}=0$ can be rewritten (for any $t\in \R$) via 
\begin{equation}
t=\frac{\textrm{cos }\al-1}{\textrm{sin }\al}
\end{equation}
as the single equation 
	\begin{equation}\label{flow}
	\cF_{\alpha \zb}=-\frac{1}{4}e^{-i
	\al}\varepsilon_{\alpha \beta \gamma}\ov{\cF}^{\beta\gamma}.
	\end{equation}
The equation \eqref{flow} is equivalent to 
\begin{equation}\label{grad1}
\boxed{
\begin{aligned}
\cF_{3\til{\gamma}}&=-e^{-i
	\al}2\veps_{\til{\gamma}}^{~\til{\alpha}}\ov{\cF}_{\til{\alpha}z}\\
	\cF_{3\zb}&=-\frac{1}{4}e^{-i
	\al}\veps^{\til{\beta}\til{\gamma}}\ov{\cF}_{\til{\beta}\til{\gamma}},
\end{aligned}}
\end{equation}
where $\til{\alpha},\til{\beta},\til{\gamma}=1,2$. These equations in fact correspond to gradient flow equations. Indeed, they can be written in the gauge $A_3=0$ (with $x^3=\tau$) as 
\begin{equation}\label{grad2}
\frac{dx^i}{d\tau}=-g^{i\ov{j}}\frac{\del \ov{W}}{\del x^{\ov{j}}}
\end{equation} 
	(using the field-space metric \eqref{fieldmetric}, and the 4d Chern-Simons functional given in \eqref{1w}),\footnote{Note that in relating \eqref{grad1} and \eqref{grad2}, as well as $\cV_0=0$ and $\mu=0$ below, one requires the condition $\phi_3=0$. This condition can be shown to be a consequence of the localization equations together with the boundary conditions on $\phi_3$, using an argument analogous to that given in Section 4.1 of \cite{analyt}.} which are gradient flow equations for a Morse function that is $2\textrm{Re}(cW)$, where $c\in \R$. This factor of $c$ is inconsequential as we are free to rescale $\tau$ in \eqref{grad2}. 
	
	Now, for $t=\pm 1$, we have $e^{i\al}=\mp i$, and 
\begin{equation}	
	W=\pm\frac{i}{g_5^2}\int_{Y\times \Sigma} dz\wedge \textrm{Tr}\bigg(\cA \wedge d\cA +{2\over 3}\cA \wedge \cA\wedge \cA\bigg).
\end{equation}	
	 Since $\textrm{Im}(W)$ is conserved along a gradient flow \cite{analyt}, $\textrm{Re}(iW)$ is conserved along the gradient flow; in fact, $\textrm{Re}(\til{c}iW)$ is conserved for any $\til{c}\in \R$. Hence, the real part of the argument of the exponent in \eqref{4dcsf} is conserved along $\til{\Gamma}$. Since the gradient flow starts from a critical point (or more precisely, a critical $H$-orbit) given by $\delta W=0$ at $x^3=\infty$ that ensures the argument of the exponent is a constant at this boundary, we find that this argument is appropriately bounded to ensure the convergence of the path integral. In addition, the boundary condition $\mu=0$  at $x^3=\infty$ ensures that the critical $H$-orbit is semistable,\footnote{It was shown that in \cite{analyt}  that only semistable $H$-orbits need to be considered to understand Stokes phenomena that occur when deforming the integration cycle.} and the fact that $\mu=0$ (that is equivalent to $\cV_0=0$), is also a localization condition is consistent with $\mu$ being conserved along gradient flows.

			Therefore, the localization equations in the form of the gradient flow equations \eqref{grad2} 
	together with the equation $\mu=0$
 define an integration cycle for 4d Chern-Simons theory that ensures its convergence. This integration cycle is the Lefschetz thimble associated with the critical point $\delta W=0$ that is a boundary condition at $x^3=\infty$. 
 

	 In order to obtain lattice models from our brane construction, we may repeat the derivation above with  fundamental strings ending on the $D4$-brane boundary at $x^3=0$. The worldlines of the endpoints of these strings realize the desired $\cQ$-invariant Wilson lines, given by 
\begin{equation}
W=\textrm{Tr}(P~e^{\int_{L }\cA_w}),
\end{equation}
where $L$ is a line along $Y\subset \del\cM$. 
In this manner, we may reproduce R-matrices, the Yang-Baxter equation with spectral parameter, and partition functions of integrable lattice models, all from a 5d partially-twisted gauge theory obtained from a type IIA configuration involving branes and fundamental strings. 
	 
	\section{Relation to 3d Chern-Simons Theory and the Geometric Langlands Correspondence} 
	In this section, we shall show how integrable lattice models realized by 4d Chern-Simons theory can be related to invariants of 3d analytically-continued Chern-Simons theory, as well as the geometric Langlands program. 
	
We shall first discuss how T-duality invariance of the partially-twisted D4-NS5 system leads to the relationship between 4d Chern-Simons theory and 3d analytically-continued Chern-Simons theory. We then modify our setup, such that the relationship can be further connected to the quantum geometric Langlands correspondence, a quantum group modification thereof, as well as a conjecture of Gaitsgory and Lurie.

\subsection{T-duality and 3d Chern-Simons Theory}
To relate to 3d analytically-continued Chern-Simons theory, we recall the D4-NS5 configuration described
in Section 3.2. Here, we further specify $\Sigma$ to be $\R\times S^1$, with $S^1$ (parametrized by $x^5$) having infinitesimal radius. Taking T-duality along this infinitesimal $S^1$ decompactifies it to $\R$.

As a result, we arrive at the following D3-NS5 configuration:
\\\vspace{0.9cm}
\begin{tikzpicture}[overlay]
\draw [decorate,decoration={brace,amplitude=4pt},xshift=74pt,yshift=-25pt]
(2.65,-0.2) -- (3.79,-0.2) node [black,midway,yshift=9pt] 
{\footnotesize $Y$};
\draw [decorate,decoration={brace,amplitude=4pt},xshift=74pt,yshift=-25pt]
(4.05,-0.2) -- (4.57,-0.2) node [black,midway,yshift=9pt] 
{\footnotesize $\mathbb{R}$};
\draw [decorate,decoration={brace,amplitude=4pt},xshift=74pt,yshift=-25pt]
(4.8,-0.2) -- (5.3,-0.2) node [black,midway,yshift=9pt] 
{\footnotesize $\mathbb{R}$};
\draw [decorate,decoration={brace,amplitude=4pt},xshift=74pt,yshift=-25pt]
(5.5,-0.2) -- (6.0,-0.2) node [black,midway,yshift=9pt] 
{\footnotesize $\mathbb{R}$};
\draw [decorate,decoration={brace,amplitude=4pt},xshift=74pt,yshift=-25pt]
(6.23,-0.2) -- (8.04,-0.2) node [black,midway,yshift=9pt] 
{\footnotesize $N\tilde{V}\!\!\subset \!\!T^\ast \tilde{V}$};
\draw [decorate,decoration={brace,amplitude=4pt},xshift=74pt,yshift=-25pt]
(2.65,0.4) -- (5.3,0.4) node [black,midway,yshift=9pt] 
{\footnotesize $\tilde{V}'$};
\draw [decorate,decoration={brace,amplitude=4pt},xshift=74pt,yshift=-25pt]
(5.5,0.4) -- (8.15,0.4) node [black,midway,yshift=9pt] 
{\footnotesize $N\tilde{V}'\!\!\subset \!\!T^\ast \tilde{V}'$};
\end{tikzpicture}
\begin{center}
\begin{tabular}{@{}lllllllllll@{}}
\toprule
   & \textbf{1} & \textbf{2} & \textbf{3} & \textbf{4} & \textbf{5} & \textbf{6} & \textbf{7} & \textbf{8} & \textbf{9} & \textbf{10}\\
   \textbf{D3} & \x & \x & \x & \x &  & & & & &\\
   $\widetilde{\textrm{\textbf{NS5}}}$ & \x & \x &  & \x & \x & \x & \x & & & \\
\bottomrule
 \end{tabular}
\end{center} 
This is a special case of the system studied by Witten \cite{witten2011fivebranes}, that realizes 3d analytically-continued Chern-Simons theory on $Y\times \R$ at $x^3=0$, with an appropriate integration cycle defined by 4d localization equations. Hence, we find that T-duality of the D4-NS5 and D3-NS5 configurations suggests a relation between 4d and 3d analytically-continued Chern-Simons theories.

This relationship can be observed at the level of twisted gauge theories. Recall from Section 3.3 that our action reduces upon dimensional reduction to that of the GL-twisted 4d super Yang-Mills action studied by Witten in \cite{witten2011fivebranes}. To see that our 5d partially-twisted theory and Witten's 4d twisted theory are in fact equivalent, we note that our theory only depends on the complex structure of $\Sigma=S^1\times \R$, and therefore we may the take the radius of $S^1\subset \Sigma$ to be infinitesimally small without changing our theory. This is because the theory only depends on the complex structure of $\Sigma$, and all cylinders share the same complex structure.
Thus, at the twisted gauge theory level, the relationship between the two theories follows from scale invariance along $\Sigma$, which amounts to rescaling of $S^1$. 

Now, in the string theory picture, if we include fundamental strings along the D4-brane boundary at $x^3=0$ to realize a lattice, they remain invariant under the operation of T-duality. As a result, we would have a lattice of Wilson lines along $Y$ in 3d analytically-continued Chern-Simons theory on $Y\times \R$. In fact, if we take $Y$ to be non-simply connected, e.g., $T^2$, this lattice forms links in $Y\times \R$, since in general the Wilson lines are located at arbitrary points along $\R$. 

In this manner, we find a relationship between lattice models realized by 4d Chern-Simons theory and link invariants of analytically-continued 3d Chern-Simons theory. It is worth noting that even without being embedded in gauge/string theory, the 3d and 4d Chern-Simons theories themselves can be shown to be T-dual 
as QFTs \cite{yamazaki}.

\subsection{S-duality and the Geometric Langlands Program}
The geometric Langlands correspondence is realized via 4d GL-twisted $\mathcal{N}=4$ SYM on a product of Riemann surfaces, $C\times (I
\times \R)$, as shown by Kapusin and Witten  \cite{kapustin2006electric}. Concisely,
shrinking the Riemann surface $C$ leads to a sigma model governing maps from $I\times \R $ into Hitchin's moduli space, and 4d S-duality gives rise to mirror symmetry of branes in the sigma model, which furnishes the geometric Langlands correspondence.

To relate our 5d theory on $Y\times \R_+ \times \R \times S^1 $ to the geometric Langlands program,\footnote{In what follows, we shall take the gauge group to be $G=SU(N)$, by freezing the center-of-mass degree of freedom of the stack of D4-branes.}  we shall use a specialized and modified version of the setup discussed in the previous section.
 We first identify $Y$ with $C$, which we take to be of genus $g>1$ (following \cite{kapustin2006electric}).\footnote{Note that, until now, we have taken $Y$ to be flat. Since the 5d topological-holomorphic theory is topological along $Y$, we can take $Y$ to be a curved manifold, via a straightforward generalization of the previous derivations. 
 } 
	Then, we replace the half-line $\R_+$ by the finite interval $I$, which also allows us to relate to the setup of 
\cite{kapustin2006electric}. Now, we ought to replace the boundary condition at $x^3=\infty$ of the half-line by a suitable boundary condition at some finite point $x^3=s$, since nonconstant gradient flows can only start at critical points if they are at infinity \cite{anewlook}. From the string theory perspective, this implies modifying the D4-NS5 system by 
another D-brane located at $x^3=s$.
We shall specify 
the relevant D-brane later, and instead first use T-duality (as described in section 5.1) to obtain a modification of Witten's setup in \cite{witten2011fivebranes} involving the D3-NS5 system. 
	Having done so, we may use the topological invariance along $C$ to shrink it to be infinitesimally small, 
	which leads us to a sigma model on $I\times \R$ with Hitchin's moduli space, $\mathcal{M}_H(G,C)$, as target space. 
	Let us understand the nature of this sigma model. We first recall that
	the twisted D3-brane worldvolume theory
	depends solely on the canonical parameter 
	\begin{equation}
	\Psi=\frac{\theta}{2\pi}+\frac{4\pi i }{g^2_{4d}}\frac{t-t^{-1}}{t+t^{-1}}.
	\end{equation}
	In addition, recall that the deformed NS5-brane boundary condition requires that 
	\begin{equation}
	\Psi=\frac{|\tau|^2}{\textrm{Re}(\tau)},
	\end{equation}
	(where $\tau$ is the complexified coupling of the worldvolume theory), i.e., $\Psi$ is real.  
	Now, following the discussion of the T-dual theory in Section 4, $t$ cannot be $\pm i $ or $\pm 1$, but obeys $|t|=1$. The restriction $t \neq \pm 1$ was necessary to obtain the full 4d Chern-Simons action, since for $t = \pm 1$ we have $w=0$. Nevertheless, since $t$ was not a relevant parameter, we could choose it to be $t=\pm 1$ in the $\cQ$-exact part of the partially twisted 5d theory, by addition of $\cQ$-exact terms to the action. Likewise, we can choose $t=\pm 1$ in the
	$\cQ_{4d}$-exact part of the  GL-twisted 4d theory, by addition of $\cQ_{4d}$-exact terms to the action (here, $\cQ_{4d}$ is the topological supercharge of the GL-twisted theory). 
	Furthermore, we can understand the real value of $\Psi$ at hand to come from a \textit{different} value of theta parameter, denoted $\theta'$, where $t=\pm 1$, i.e.,
	\begin{equation}\label{can}
\Psi=\frac{\theta'}{2\pi}=\textrm{Re }\tau'.
\end{equation}
For $t=\pm 1$, we know that the corresponding sigma model is in fact an A-model in a symplectic structure (of the target space) proportional to $\omega_K$ (following the convention of \cite{kapustin2006electric}), 
	 and we shall employ this description of the sigma model in what follows.\footnote{In fact, it can be shown that for all $t$ satisfying $|t|=1$ with $t\neq \pm i$, the sigma model is a $B$-field transform of an A-model with symplectic structure proportional to $\omega_K$.}


	Now, the NS5-boundary condition we used on one end of the interval can be interpreted as a space-filling coisotropic brane, $\cB_c$, of the sigma model, of type $(B,A,A)$ . To see this, first note that the relevant fields $A_1$, $A_2$, $\phi_1$ and $\phi_2$ that enter the sigma model obey generalized Neumann boundary conditions in the twisted 4d worldvolume theory, which correspond to a space-filling sigma model brane. In addition, recall that the sigma model includes an action for the $B$-field 
	\begin{equation}
	B=-\frac{\theta'}{2\pi}\omega_I,
	\end{equation}
	where 
	\begin{equation}\label{wi}
	\omega_I=-\frac{1}{4\pi}\int_C \textrm{Tr} (\delta A\wedge \delta A-\delta \phi\wedge \delta \phi). 
	\end{equation}
	This $B$-field is nondegenerate on the brane, and satisfies $(\omega^{-1}B)^2=-1$, where $\omega=\Psi \omega_K$ is the defining symplectic structure of the A-model, meaning that the brane is indeed an A-brane (that is coisotropic) with respect to $\omega$. Moreover, the $B$-field is of type (1,1) in complex structure $I$, which implies that the brane is a B-brane in this complex structure.
Finally, in the localization limit whereby the $\cQ$-exact term of the sigma model is multiplied by a large factor, the boundary conditions reduce to the boundary restriction of the localization equations (i.e., holomorphic maps), as expected for a coisotropic A-brane.	
	
	Next, at $x^3=s$, we shall pick a boundary condition that will eventually enable us to realize the quantum geometric Langlands correspondence. This boundary condition is that of an ($A,B,A$) brane, $\cB'$, that is Lagrangian with respect to the symplectic form \eqref{wi}. In particular, such a brane arises when $\cM_H(G,C)$ is described (using the complex structure $J$) as the moduli space of semi-stable flat $G_{\C}$-bundles on $C$. In this description, an open subset of $\cM_H(G,C)$ maps to $\cM(G,C)$, the moduli space of semi-stable $G$-bundles, and the fibers of the map are the ($A,B,A$) branes of interest. From the perspective of type IIB string theory, these sigma model branes arise from D5-branes \cite{gf}.

	In fact, we have obtained the necessary ingredients to realize the structure of a twisted D-module, due to the presence of the coisotropic brane, $\mathcal{B}_{c}$ at $x^3=0$, as well as the Lagrangian property of $\cB'$. This follows since we can also define the sigma model with $\mathcal{B}_{c}$ boundary conditions at both $x^3=0$ and $x^3=s$, and the sheaf of ($\mathcal{B}_{c},\mathcal{B}_{c}$) strings can be shown to be the sheaf $\cD_{K^{1/2}_{\cM}\otimes \mathcal{L}^{\Psi}}$ of holomorphic differential operators acting on sections of ${K^{1/2}_{\cM}\otimes \mathcal{L}^{\Psi}}$, where $\cM=\cM(G,C)$ and where $\mathcal{L}$ is the determinant line bundle on $\cM(G,C)$ \cite{kapustin2006electric}. Moreover, the ($\mathcal{B}_{c},\mathcal{B}'$) strings can be shown to be sections of a tensor product bundle that includes ${K^{1/2}_{\cM}\otimes \mathcal{L}^{\Psi}}$.
	
	
	
	Now, under type IIB S-duality, which results in S-duality of the worldvolume theory, the canonical parameter $\Psi$ of the D3-NS5 brane system transforms to $^L\Psi=-\frac{1}{\Psi}$. In addition, since $t$ is not a relevant parameter, it can be considered to be invariant under S-duality. 
This implies that the effective 2d A-model with $\cM_H(G,C)$ as target space is dualized to another A-model in symplectic structure $\omega_K$ 
of the target $\cM_H(^LG,C)$, i.e., Hitchin's moduli space for the Langlands dual gauge group $^LG$. Moreover, under this duality, ($B,A,A$) and ($A,B,A$) branes in $\cM_H(G,C)$ are mapped respectively to ($B,A,A$) and ($A,B,A$) branes in $\cM_H(^LG,C)$. This in fact gives rise to a realization of the quantum geometric Langlands correspondence that maps twisted D-modules to twisted D-modules similar to that described in Section 11.3 of \cite{kapustin2006electric}.

To understand why the $B_c$ brane on $\cM_H(G,C)$ maps to the $B_c$ brane on $\cM_H(^LG,C)$  under S-duality from the perspective of type IIB string theory, first note that the duality of sigma models above can be generalized such that the canonical parameter transforms to $^L\Psi=\frac{1}{\Psi}$ instead. This happens when we use the orientation reversal symmetry of the 4d worldvolume theory  prior to taking S-duality, which results in $\Psi\rightarrow - \Psi$. 

Also recall that $B_c$ originated from the deformed NS5-brane boundary condition of the D3-brane worldvolume theory. This deformation arises from the nonzero value of $\theta$, which is the expectation value of a Ramond-Ramond scalar.  Now, under the $SL(2,\Z)$ transformation that maps the complex coupling parameter $\tau$ to $\tau + 1$, $\theta$ shifts to $\theta+2\pi$. 
This shift converts the NS5-brane to a (1,1) fivebrane, while the canonical parameter 
of the GL-twisted theory is shifted as ${\Psi}\rightarrow\Psi+1$. Since the effect is just a shift of the $\theta$ parameter, the sigma model description of the boundary condition as a $B_c$ brane is unaffected, and we can just consider this to be our starting point where the canonical parameter is $\Psi$.

Now, orientation reversal symmetry further converts this (1,1) fivebrane to a (1,-1) fivebrane. Then, under the $SL(2,\Z)$ transformation that maps 
the gauge group $G$ to $^LG$ and $\tau$ to $-\frac{1}{\tau}$, 
this (1,-1) fivebrane becomes a (1,1) fivebrane, and the resulting canonical parameter is $\frac{1}{\Psi}$. As before, the latter gives rise to generalized Neumann boundary conditions for the relevant fields $A_1$, $A_2$, $\phi_1$ and $\phi_2$, but now in the S-dual 4d gauge theory. In the aforementioned localization limit of the effective sigma model, these reduce to the boundary restriction of the localization equations, just as in the original theory with gauge group $G$. This implies that the corresponding sigma model brane is the coisotropic brane $B_c$ on $\cM_H(^LG,C)$.
 \newline
 \newline
 	\noindent{\textit{S-duality of 3d Analytically-continued Chern-Simons Theory}}

Now, we have a duality of the 4d D3-brane worldvolume theory that maps the gauge group $G$ to its Langlands dual $^LG$ and (1,1) fivebranes to themselves, and moreover both systems localize to 3d analytically-continued Chern-Simons theory at the boundary.
This can be interpreted as an S-duality for 3d analytically-continued Chern-Simons theory, which has been predicted previously in \cite{TY,DimofteGukov}. Note that the aforementioned $SL(2,\Z)$ transformation that converts the NS5-brane to a (1,1) fivebrane also converts fundamental strings to (1,1) strings, and therefore the Wilson lines that realize links (as discussed in Section 5.1) ought to become Wilson-'t Hooft lines  at $x^3=0$ in the 4d worldvolume theory, and therefore Wilson-'t Hooft lines in the equivalent 3d analytically-continued Chern-Simons theory. Moreover, the duality of the 4d worldvolume theory that gives rise to S-duality of Chern-Simons theory maps (1,1) strings to themselves, implying that Wilson-'t Hooft lines for the gauge group $G$ are mapped to Wilson-'t Hooft lines for the gauge group $^LG$. Note that although the full $G_{\C}/^LG_{\C}$ symmetry of the S-dual Chern-Simons theories does not enter the definition of the Wilson-'t Hooft lines, they are nevertheless expected to exist from consideration of supersymmetry of the twisted 4d theory at $x^3=0$. 
\subsection{A Modification of the Quantum Geometric Langlands Correspondence}
If we include the fundamental strings that realize lattices and links as previously discussed, the quantum geometric Langlands correspondence is actually generalized in our case, as follows. Recall that to realize integrable lattice models using 4d Chern-Simons theory, we ought to include a network of Wilson lines along $x^3=0$ in our setup above. Since $\Sigma=\R\times S^1$, these Wilson lines are in fact classified by representations of a quantum affine algebra \cite{CWY2}.

These Wilson lines are located at points on $\Sigma=\R\times S^1$, and upon shrinking $S^1$ (using the holomorphic invariance of the partially-twisted D4-brane worldvolume theory), they are located on points along $\R$, and are now determined by representations of a quantum group 
that descend from representations of the aforementioned quantum affine algebra \cite{CWY2}. Then, the subsequent $SL(2,\Z)$ transformation that maps the NS5-brane to a (1,1) fivebrane maps fundamental strings to (1,1) strings, and therefore Wilson lines to Wilson-'t Hooft lines.

Upon shrinking $Y$, they then become local operators located at $x^3=0$ and along $\R$. Therefore, they correspond to points on the coisotropic brane.
Hence, this modifies the twisted D-modules 
that appear in quantum geometric Langlands to involve the data of the Wilson-'t Hooft lines (labelled by an index, $i$), namely, homomorphisms $\rho^i:U(1)\rightarrow G$ ('t Hooft) and representations of the quantum deformation of a subgroup of $G$ (Wilson). This subgroup, denoted $G_{\rho^i}$, is that which commutes with $\rho^i(U(1))$.\footnote{The definition of an 't Hooft line operator for the gauge group, $G$, involves embedding a Dirac monopole into $G$, via a homomorphism $\rho:U(1)\rightarrow G$, and this continues to be true for a Wilson-'t Hooft line operator. In the presence of the latter, the gauge field is $\rho(A_0)+\hat{A}$, where $A_0$ is the singular $U(1)$ gauge field with Dirac singularity, and $\hat{A}$ is smooth. To ensure that the singular part of the curvature coincides with the Dirac singularity and does not depend on $\hat{A}$, we need $[\rho(A_0),\hat{A}]=0$. However, this results in gauge transformations along the operator being restricted to the subgroup of $G$ that commutes with $\rho(U(1))$.}  


Hence, this indicates that in our present setup that realizes lattices/links,
quantum geometric Langlands is actually modified by the data of quantum group representations. In particular, the dual categories of twisted D-modules on $\cM(G,C)$ and $\cM(^LG,C)$ are modified by representations of quantum deformations of $G_{\rho^i}$ and $^LG_{^L\rho^i}$, respectively, where $^L\rho^i:U(1)\rightarrow\textrm{} ^LG$. 

\subsection{The Gaitsgory-Lurie Conjecture}

The relationship between 3d analytically-continued Chern-Simons theory and the quantum geometric Langlands correspondence 
also realizes 
a known extension 
of this correspondence 
to quantum groups, conjectured by Gaitsgory and Lurie \cite{Gaitsgory}. Here, the correspondence is between the Kazhdan-Luzstig category, $\textrm{KL}_{\Psi}(G)$, of finitely generated modules over the affine Kac-Moody algebra $\widehat{\mathfrak{g}}_{\Psi-h^{\vee}}$ (where $h^{\vee}$ is the dual Coxeter number of $G$) on which the action of $\mathfrak{g}[[z]]$ integrates to an action of the group $G[[z]]$, and the category of Whittaker D-modules on the affine Grassmannian $\textrm{Gr}(^LG)$ of $^LG$, denoted $\textrm{Whit}_{\frac{1}{\Psi}}(^LG)$.\footnote{The Whittaker category $\textrm{Whit}_{c}$ consists of $c$-twisted D-modules on Gr$_{G}$ that are $N((z))$-equivariant with respect to a non-degenerate character, where $N$ is the maximal unipotent subgroup of $G$.} The former category is in fact equivalent to the category of representations of quantum groups $U_q(G)$, where $q$ is related to $\Psi$ via $q=\textrm{exp}(\frac{\pi i}{\Psi})$, with $\Psi=k+h^{\vee}$ (where $k$ is the level of the affine Kac-Moody algebra).
 \newline
 \newline
 	\noindent{\textit{Realizing the Gaitsgory-Lurie Conjecture}}

In the present setup, the conjectured correspondence can be realized as follows.
The deformed NS5-brane boundary condition leads to 3d analytically-continued Chern-Simons theory on the boundary $C\times \R$, where Wilson lines (realized by fundamental strings) are admissible as operators, and form links. 
These Wilson lines are labelled by representations of the quantum deformation of $G_\C$ that descend from representations of the corresponding quantum affine algebra \cite{CWY2} when we shrink $S^1$, as described in Section 5.1. Now, we may deform the integration cycle of 3d analytically-continued Chern-Simons theory to one that parametrizes real gauge fields, and upon making the identification $\Psi=k+h^{\vee}$, we obtain ordinary 3d Chern-Simons theory \cite{witten2011fivebranes}, where $k\in \Z_+$ is the level (more generally, if we also allow $k\in \Z_-$, we require $\Psi=k+\textrm{sign}(k)h^{\vee}$).\footnote{The possibility of such a deformation does not imply an S-duality for real 3d Chern-Simons theory, since in general $\frac{1}{k+h^{\vee}}$ is not an integer.} Upon doing so, the Wilson lines will be in representations of $U_q(G)$, where $q=\textrm{exp}(\frac{\pi i}{k+h^\vee})$.


Moreover, the Wilson lines can be taken to lie along $\R$ and on points on $C$, by deforming the links appropriately using topological invariance of Chern-Simons theory. As a result, there are local operators of the corresponding WZW theory on $C$ (at any point on $\R$) associated with the quantum group representations, giving rise to the aforementioned finitely generated module over $\widehat{\mathfrak{g}}_{k}$, denoted $\widehat{\mathfrak{g}}_{\Psi-h^{\vee}}\textrm{-}mod_C^0$.

Under an orientation reversal that sends $\Psi\rightarrow -\Psi$ and S-duality that sends $-\Psi \rightarrow \frac{1}{\Psi}$, the NS5-brane becomes a D5-brane that realizes the maximal Nahm pole boundary condition on the $\phi_1$, $\phi_2$ and $\phi_4$ fields \cite{GW}. Moreover, the Wilson lines realized by fundamental strings become 't Hooft lines realized by D1-branes. The category of these 't Hooft lines is precisely the Whittaker category, as explained in \cite{gf}.
 \newline
 \newline
 	\noindent{\textit{Realizing the Vertex Algebra Version of the Gaitsgory-Lurie Conjecture}}

We can also realize the vertex algebra version of the Gaitsgory-Lurie conjecture, as studied by Aganagic, Frenkel, and Okounkov \cite{aga}. In this version, the Whittaker category is replaced by a certain subcategory 
of the category of modules of the affine W-algebra, $\mathcal{W}_{\frac{1}{\Psi}}(^L\mathfrak{g})$, whose objects are modules, denoted $\mathcal{W}_{\frac{1}{\Psi}}(^L\mathfrak{g})\textrm{-}mod^0$, that correspond to ``magnetic"  vertex operators. We can understand how this arises physically on $C$.
Firstly, the D5-brane boundary condition was shown by Gaiotto and Witten \cite{GWK} to have a 
description in terms of $^L\mathfrak{g}$ opers (with singularities due to the 't Hooft lines), which describe $\mathcal{W}_{\frac{1}{\Psi}}(^L\mathfrak{g})$ conformal blocks on $C$ in the classical limit. 
Furthermore, with 't Hooft line knots/links, braiding along $\R$ of the $\R$-independent \textit{quantum} BPS states of the D3-D5 system was identified with braiding of complex integration cycles for 3d analytically-continued Chern-Simons theory on $C\times I$ with coupling $\frac{1}{\Psi}$. This in turn is related via deformation of its complex integration cycle to one that parametrizes real gauge fields, as well as the oper boundary condition, to 
$\mathcal{W}_{\frac{1}{\Psi}}(^L\mathfrak{g})$ conformal blocks on $C$. Hence, the vertex algebra form of the Gaitsgory-Lurie conjecture ought to be realizable as the equivalence of quantum BPS states under S-duality.
 \newline
 \newline

 	\noindent{\textit{Realizing the Quantum $q$-Langlands Correspondence}}

In the present setup, one should also be able to realize the quantum $q$-Langlands correspondence \cite{aga} that relates modules of the quantum deformation of the affine algebra, $\widehat{\mathfrak{g_\C}}$, and the $q$-deformed affine W-algebra for $^L\mathfrak{g_\C}$, by T-dualizing both the D3-NS5 and D3-D5 systems. The former leads to the D4-NS5 system studied in previous sections that gave us 4d Chern-Simons theory on $C\times \R \times S^1$, with Wilson lines in representations of the quantum deformation of 
$\widehat{\mathfrak{g_\C}}$. On the other hand, T-dualizing the D3-D5 system  in an appropriate direction gives us a D4-D6 system, while the D1-branes realizing 't Hooft lines become D2-branes realizing 't Hooft surface operators along $\R \times S^1$ and on points on $C$. Given the known T-duality between 3d and 4d Chern-Simons theories, as well as the fact that the D4-D6 system also realizes the Nahm pole boundary condition, it is reasonable to expect that we can find a generalization of Gaiotto-Witten's description of the D3-D5 BPS states. That is, we expect that braiding along $\R$ of the $\R$-independent quantum BPS states of the D4-D6 system can be identified with braiding of complex integration cycles for 4d Chern-Simons theory on $C\times I \times S^1$. This in turn ought to be related via the Nahm pole boundary condition to a 3d boundary theory on $C\times S^1$ that realizes modules of the $q$-deformed affine W-algebra for $^L\mathfrak{g_\C}$. 
 \newline
 \newline
 	\noindent{\textit{Summary}}

The results above are summarized in Figure 2.
 \begin{figure}
 \catcode`\@=11
\newdimen\cdsep
\cdsep=3em

\def\cdstrut{\vrule height .6\cdsep width 0pt depth .4\cdsep}
\def\@cdstrut{{\advance\cdsep by 2em\cdstrut}}

\def\arrow#1#2{
  \ifx d#1
    \llap{$\scriptstyle#2$}\left\downarrow\cdstrut\right.\@cdstrut\fi
  \ifx u#1
    \llap{$\scriptstyle#2$}\left\uparrow\cdstrut\right.\@cdstrut\fi
  \ifx r#1
    \mathop{\hbox to \cdsep{\rightarrowfill}}\limits^{#2}\fi
  \ifx l#1
    \mathop{\hbox to \cdsep{\leftarrowfill}}\limits^{#2}\fi
}
\catcode`\@=12

\cdsep=3em
$$
\begin{matrix}
 & & \textrm{KL}_{\Psi}(G)/\widehat{\mathfrak{g}}_{\Psi-h^{\vee}}\textrm{-}mod_C^0                & \arrow{r}{\cS'}   & \textrm{Whit}_{\frac{1}{\Psi}}(^LG)/\mathcal{W}_{\frac{1}{\Psi}}(^L\mathfrak{g})\textrm{-}mod^0_C                  \cr
 &&   \arrow{u}{\cT^{-1}} &                      & \arrow{u}{\cS'\cT^{-1}\cS'^{-1}} \cr
&  \textrm{4d CS}_{\Psit }(G_{\C})                    & \arrow{r}{\cT \textrm{T}}            \textrm{3d CS}_{\Psi +1}(G_{\C})  & \arrow{r}{\cS'}   & \textrm{3d CS}_{\frac{1}{\Psi +1} }(^LG_{\C})                    \cr
 && \arrow{d}{C\rightarrow 0} &                      & \arrow{d}{C\rightarrow 0} \cr
&&  D\textrm{-}mod^{U_{q'}(G_{\rho^i}),\rho^i}_{\Psi+1}(\cM(G,C))                 & \arrow{r}{\cS'} & D\textrm{-}mod^{U_{^Lq'}(^LG_{^L\rho^i}),^L\rho^i}_{\frac{1}{\Psi+1}}(\cM(^LG,C))                  \cr
\end{matrix}
$$ 
\caption{A relationship between 4d Chern-Simons theory, 3d S-dual Chern-Simons theories, the quantum group modification of quantum geometric Langlands, and the Gaitsgory-Lurie conjecture. Here, $q'=\textrm{exp}(\frac{\pi i}{\Psi+1})$.
}
\end{figure}
Here, T denotes T-duality from type IIA string theory to type IIB string theory, $\cS'=\cS R$ where $R$ is the orientation reversal symmetry transformation of the twisted D3-brane worldvolume theory, and where $\cS$ and $\cT$ are the generators of the $SL(2,\Z)$ symmetry of type IIB string theory.
\section{Conclusion and Future Work}
In this work, we have shown that integrable lattice models, link invariants, quantum geometric Langlands, and the Gaitsgory-Lurie conjecture are related via dualities in string theory. These dualities manifest as invariances in the relevant spectra of the respective worldvolume theories. The crucial ingredient is the fact that the 5d $\cN=2$ SYM theory of a stack of D4-branes admits a partial twist that is topological-holomorphic, and analogous to the GL-twist of 4d $\cN=4$ SYM.

This 
suggests generalizations of our present work. Firstly, we should be able to define GL type twists for maximally supersymmetric Yang-Mills theory in 6d and 7d as well, which would lead to 5d and 6d Chern-Simons theories once NS5-type boundary conditions are imposed. Then, we would find that all Chern-Simons theories in dimensions three to six are related via a chain of T-dualities that relate the D3-NS5, D4-NS5, D5-NS5 and D6-NS5 systems, in accordance with their known T-duality at the level of field theory \cite{yamazaki}.

Moreover, further T-dualities applied to the S-dual D3-D5 system ought to furnish  higher analogues of the quantum geometric Langlands correspondence, just as the D4-D6 system ought to realize the quantum $q$-Langlands correspondence via TST-duality to the D4-NS5 system. Namely, we expect the quantum $q,\textrm{}v$-geometric Langlands correspondence involving elliptic affine W-algebras to follow from T-duality to the D5-D7 system, and a further generalization to be furnished by T-duality to the D6-D8 system.


We hope to explore these issues in future work.
\appendix
\label{App:GammaMat}
\section{5d Gamma Matrices and Spinor Operations} 

Where necessary in Section 2, we use the following 
representation of the gamma matrices in five (Euclidean) dimensions
\begin{align}\label{eq:GammaMatrices}
\Gamma^1=\si^1\otimes\si^3, && \Gamma^2=\si^2\otimes\si^3, && \Gamma^3=\si^3\otimes\si^3 && \Gamma^4=\mathds{1}\otimes\si^1, && \Gamma^5=\mathds{1}\otimes\si^2, 
\end{align}
where the $\{\si^1,\si^2,\si^3\}$ are the Pauli matrices, i.e.,
\begin{align}
\si^1=\left(
\begin{array}{cc}
0 & 1 \\
1 & 0
\end{array}\right),
&&
\si^2=\left(
\begin{array}{cc}
0 & -i \\
i & 0
\end{array}\right),
&&
\si^3=\left(
\begin{array}{cc}
1 & 0 \\
0 & -1
\end{array}\right).
\end{align}
These gamma matrices obey the Clifford algebra
\begin{equation} \label{gamma}
\left\{\Gamma_M,\Gamma_N\right\}=2g_{MN}\mathds{1}_{4\times 4}.
\end{equation}
In addition, this set of gamma matrices is also used for the R-symmetry group $SO(5)_R$.

The 
$SO(5)$ rotation/R-symmetry group spinor indices are lowered and raised using the two index antisymmetric tensor $\Omega$, i.e.,\footnote{We shall only write formulas corresponding to rotation group spinors in what follows; the corresponding formulas for R-symmetry group spinors can be obtained by replacing indices with hatted versions of themselves.}
\begin{align}
\rho_A=\rho^B\Omega_{BA}, && \rho^A=\Omega^{AB}\rho_B, \end{align}
where $\rho_A$ and $\rho^A$ correspond to the representation $\bf{4}$ and its dual representation $\bf{4}^{\vee}$.
Here, $\Omega$ is 
\begin{equation}
\Omega^{AB}=\eps^{\balpha\bbeta}\otimes B^{\bm\bn}=\begin{pmatrix} 0 & 1 \\ -1 & 0 \end{pmatrix}\otimes \begin{pmatrix} 0 & 1 \\ 1 & 0 \end{pmatrix}.
\end{equation}
Furthermore, the two index antisymmetric tensor $\eps$ can be used to lower and raise 
$SO(3)$ spinor indices, i.e.,
\begin{align}
\lambda_{\balpha}=\lambda^{\bbeta}\eps_{\bbeta\balpha}, && \lambda^{\balpha}=\eps^{\balpha\bbeta}\lambda_{\bbeta}. \end{align}
In particular, this antisymmetric tensor acts on the Pauli matrices to give symmetric matrices, i.e.,  $(\si^{\alpha})_{\balpha}^{~~\bbeta}\eps_{\bbeta\bgamma}=(\si^{\alpha})_{\balpha\bgamma}$  and $\eps^{\balpha\bbeta}(\si^{\alpha})_{\bbeta}^{~~\bgamma}=(\si^{\alpha})^{\balpha\bgamma}$, where $(\si^{\alpha})_{\balpha\bgamma}=(\si^{\alpha})_{\bgamma\balpha}$ and $(\si^{\alpha})^{\balpha\bgamma}=(\si^{\alpha})^{\bgamma\balpha}$.

\section{Derivation of $\cQ$-invariant Boundary Term of 5d Partially-Twisted Theory}
Here, we explain the derivation of \eqref{long} from the $t$-dependent terms \eqref{s2} and \eqref{liftedboundaryint} of the 5d partially-twisted 
worldvolume theory of a stack of D4-branes ending on a (deformed) NS5-brane. Explicitly, these $t$-dependent terms are 
\begin{equation}\label{tdep}
	\begin{aligned}
	S_3+S_{\del \cM}=&-{1\over {g_5}^2}\int_\cM d^5x\ \varepsilon^{\alpha\beta\gamma}\tr\left(2\cbrac{{t-t^{-1}\over t+t^{-1}}}\cbrac{\half F_{\alpha 4}F_{\beta\gamma}+\half\partial_\alpha\cbrac{\phi_\beta D_4\phi_\gamma}+\partial_\alpha\cbrac{F_{\beta 5}\phi_\gamma}}\right.\\
	&\qquad\qquad\qquad\qquad\qquad\left.\phantom{\half}-\cbrac{{4\over t+t^{-1}}}\cbrac{\half F_{\alpha 5}F_{\beta\gamma}+\half\partial_\alpha\cbrac{\phi_\beta D_5\phi_\gamma}-\partial_\alpha\cbrac{F_{\beta 4}\phi_\gamma}}\right)\\
	&+{1\over {g_5}^2}\int_{\partial \mathcal{M}} d^4x\ \tr\left(\cbrac{t+t^{-1}}\cbrac{\half\varepsilon^{{\tal}{\tbt} }D_5\phi_{\tal}\phi_{\tbt} }\right.\\
	&\qquad\qquad\qquad\qquad\left.\phantom{\half}+\cbrac{{t+t^{-1}\over t-t^{-1}}}\varepsilon^{ijk}\cbrac{A_i\partial_j A_k+{2\over 3}A_i A_j A_k}\right).
	\end{aligned}
	\end{equation}
	Now, note that 
	\begin{equation}
	\textrm{Tr}(\veps^{\al\bt\gm4}F_{\al\bt}F_{\gm 4})=\textrm{Tr}\bigg(\frac{1}{4}\veps^{\mu\nu\rho\si}F_{\mu\nu}F_{\rho\si}\bigg)
	\end{equation}
	for $\mu,\nu,\rho,\si= 1,2,3,4$ where the Levi-Civita symbol is defined such that $\veps^{1234}=1$, and analogously
		\begin{equation}
	\textrm{Tr}(\veps^{\al\bt\gm5}F_{\al\bt}F_{\gm 5})=\textrm{Tr}\bigg(\frac{1}{4}\veps^{\til{\mu}\til{\nu}\til{\rho}\til{\si}}F_{\til{\mu}\til{\nu}}F_{\til{\rho}\til{\si}}\bigg)
	\end{equation}
	for $\til{\mu},\til{\nu},\til{\rho},\til{\si}= 1,2,3,5$ and $\veps^{1235}=1$. Moreover, these expressions can be written as total derivatives using
	\begin{equation}
	\textrm{Tr}\bigg(\frac{1}{4}\veps^{\mu\nu\rho\si}F_{\mu\nu}F_{\rho\si}\bigg)=\veps^{\mu\nu\rho\si}\del_{\mu}\textrm{Tr}\bigg(A_{\nu}\del_{\rho}A_{\si}+\frac{2}{3}A_{\nu}A_{\rho}A_{\si}\bigg)
	\end{equation}
	and 
		\begin{equation}
	\textrm{Tr}\bigg(\frac{1}{4}\veps^{\til{\mu}\til{\nu}\til{\rho}\til{\si}}F_{\til{\mu}\til{\nu}}F_{\til{\rho}\til{\si}}\bigg)=\veps^{\til{\mu}\til{\nu}\til{\rho}\til{\si}}\del_{\til{\mu}}\textrm{Tr}\bigg(A_{\nu'}\del_{\rho}A_{\til{\si}}+\frac{2}{3}A_{\til{\nu}}A_{\til{\rho}}A_{\til{\si}}\bigg).
	\end{equation}
	
	Hence, all the terms in $S_3$ can be written as total derivatives, and using Stoke's theorem, we find that the entire $t$-dependent action \eqref{tdep} can be written as the boundary action 
\begin{equation}\label{bapp}
\begin{aligned}
\frac{1}{g_5^2}\int_{\del\cM}d^4x\textrm{Tr}\Bigg(&
T\varepsilon^{ijk}\cbrac{A_i\partial_j A_k+{2\over 3}A_i A_j A_k}+Tw\bigg(2\veps^{\tal\tbt}F_{\tbt 4}\phi_{\tal}\bigg)-Tw^2\veps^{\tal\tbt}\phi_{\tal}D_4\phi_{\tbt}
\\&+Tw\varepsilon^{\til{i}\til{j}\til{k}}\cbrac{A_{\til{i}}\partial_{\til{j}} A_{\til{k}}+{2\over 3}A_{\til{i}} A_{\til{j}} A_{\til{k}}}+Tw^2\bigg(2\veps^{\tal\tbt}F_{\tbt 5}\phi_{\tal}\bigg)-Tw^3\veps^{\tal\tbt}\phi_{\tal}D_5\phi_{\tbt},
\Bigg)
\end{aligned}
\end{equation}	
where $\til{i}, \til{j},\til{k}=1,2,5$, and 
\begin{equation}
T=-\frac{t-t^{-1}}{t+t^{-1}}+\frac{t+t^{-1}}{t-t^{-1}}.
\end{equation}  
Here, we have used 
\begin{equation}
Tw=\frac{2}{t+t^{-1}},
\end{equation}  
\begin{equation}
Tw^2=\frac{t-t^{-1}}{t+t^{-1}},
\end{equation}  
and
\begin{equation}
Tw^3=\frac{t+t^{-1}}{2}-\frac{2}{t+t^{-1}}.
\end{equation}  
The boundary action \eqref{bapp} is equal to \eqref{long}. This follows from $\Psit=-\frac{4\pi i}{g_5^2}T$, and can be checked by expanding \eqref{long} in terms of $A_1$, $A_2$, $\phi_1$ and $\phi_2$.
	
\end{document}